\documentclass[aps,prd,amsmath,floats,floatfix,
  superscriptaddress,
  nofootinbib,
  notitlepage,
  showpacs]{revtex4-1}
  
\usepackage{ulem}
\usepackage{sidecap}
\usepackage{amssymb}
\usepackage{amsmath}
\usepackage{verbatim}
\usepackage{mathrsfs}
\usepackage{amsfonts}
\usepackage{latexsym}
\usepackage{epsfig}
\usepackage{color}
\usepackage{graphicx}
\usepackage{units}
\usepackage{overpic}
\usepackage{afterpage}
\usepackage{float}
\usepackage{mathtools}
\usepackage{xcolor}
\usepackage{ulem}
\usepackage[hidelinks]{hyperref}
\hypersetup{
    colorlinks,
    linkcolor={red!50!black},
    citecolor={blue!50!black},
    urlcolor={blue!80!black}
}

\usepackage{booktabs, multirow} % for borders and merged ranges
\usepackage{soul}% for underlines
\usepackage{changepage,threeparttable} % for wide tables
\begin{document}

%%%%%%%%%%%%%%%%%%
%%%   MACROS   %%%
%%%%%%%%%%%%%%%%%%

\definecolor{orange}{rgb}{0.9,0.45,0} 
\definecolor{applegreen}{rgb}{0.055, 0.591, 0.0530}
\newcommand{\vjp}[1]{{\textcolor{blue}{[VJ: #1]}}}
\newcommand{\nico}[1]{{\textcolor{brown}{[nico: #1]}}}
\newcommand{\dario}[1]{{\textcolor{red}{[Dario: #1]}}}
\newcommand{\argelia}[1]{{\textcolor{orange}{[Argelia: #1]}}}

%%%%%%%%%%%%%%%%%%%%%%%%%%%%%
%%%   TITLE AND AUTHORS   %%%
%%%%%%%%%%%%%%%%%%%%%%%%%%%%%

\title{
Gravitational confinement of ghost scalar fields in neutron stars}

\author{Argelia Bernal}%\orcidlink{0000-0003-0295-0053}
\affiliation{Departamento de F\'isica, Divisi\'on de Ciencias e Ingenier\'ias, Campus Le\'on, Universidad de Guanajuato, C.P. 37150, Le\'on, M\'exico.}

\author{V\'ictor Jaramillo}%\orcidlink{0000-0002-3235-4562} 
\affiliation{Department of Modern Physics, University of Science and Technology of China, Hefei, Anhui 230026, China}
\affiliation{Departamento de Física, Universidad de Guadalajara, 44430 Guadalajara, Jalisco, México}

\author{N\'estor A. Montiel-Hern\'andez}%\orcidlink{0000-0003-0295-0053}
\affiliation{Departamento de F\'isica, Divisi\'on de Ciencias e Ingenier\'ias, Campus Le\'on, Universidad de Guanajuato, C.P. 37150, Le\'on, M\'exico.}

\author{Dar\'io N\'u\~nez}%\orcidlink{0000-0003-0295-0053}
\affiliation{Instituto de Ciencias Nucleares, Universidad Nacional
  Aut\'onoma de M\'exico, Circuito Exterior C.U., A.P. 70-543,
  Coyoac\'an, M\'exico 04510, CdMx, M\'exico}

\author{Nicolas Sanchis-Gual}
\affiliation{Departament d'Astronomia i Astrof\'isica, Universitat de Val\`encia, Av. Vicent Andr\'es Estell\'es 19, 46100 Burjassot (Val\`encia), Spain}

\begin{abstract}
We investigate the effects, stability, and nonlinear dynamics of ghost scalar matter modeled as a field with a negative kinetic term confined within the cores of neutron stars. To this end, we analyze static configurations of the coupled Einstein–Euler–(ghost, complex) Klein–Gordon system and then we perform fully dynamical numerical evolutions of illustrative cases. Our results demonstrate that neutron stars can gravitationally confine a finite amount of ghost matter and support continuous families of equilibrium solutions, indicating that these configurations are not the result of fine tuning. We analyze the properties of the final states and find that the neutron star undergoes a persistent pulse-like oscillatory motion. In particular, we explicitly compute the frequency synchronization between the stellar fluid oscillation modes and those of the ghost scalar sector.
\end{abstract}

\date{\today}

\maketitle

\section{Introduction}

Ghost matter arises as a controversial concept, since a negative kinetic term in the Lagrangian challenges the very notion of positive-definite energy and, at the quantum level, leads to severe instabilities. Nevertheless, even at the classical level it poses interesting conceptual questions: (i) what is the fate of a system whose Hamiltonian is not bounded from below, and (ii) how should one interpret Lagrangian and Hamiltonian formulations when the kinetic structure departs from standard assumptions?

Beyond these foundational issues, matter sources that violate the standard energy conditions frequently appear in gravitational physics. Several theoretical alternatives to black holes, as well as proposals for exotic compact objects in extended theories of gravity, require matter with unconventional properties~\cite{cardoso2019testing}. A notable example is the gravastar proposal~\cite{mazur2004gravitational,visser2004stable,bronnikov2022black,jampolski2024nested}, in which an internal de Sitter region is matched to an external Schwarzschild geometry through a shell of highly exotic matter, thereby avoiding the formation of an event horizon. Such configurations cannot be supported by ordinary matter alone. At a minimum, violations of the strong energy condition are required, and in most stable realizations the null and weak energy conditions are violated as well. Traversable wormholes provide another well-known example~\cite{visser1995lorentzian,lemos2008black}, where violation of the null energy condition is unavoidable within classical General Relativity. Finally, the observed accelerated expansion of the Universe is compatible with matter sources that violate the strong energy condition. In particular, models with an effective equation of state parameter $w<-1$, often referred to as phantom dark energy, correspond to violations of the null energy condition and can be modeled, at least phenomenologically, by scalar fields with negative kinetic terms, see \cite{Vazquez:2023kyx} for instance.

Ghost scalar fields have been extensively investigated in cosmological settings~\cite{Najera:2024phantom} and in the construction of wormhole geometries. In the latter context, several works have analyzed the stability properties of $3+1$-dimensional scalar-field wormholes within General Relativity (see, e.g., \cite{Matos:2005uh,Gonzalez:2008wd,Shinkai:2002gv,Gonzalez:2008xk,Dzhunushaliev:2008bq}). These studies include, among others, the massless Bronnikov (Ellis) wormhole~\cite{Bronnikov:1973fh} as well as configurations supported by self-interacting scalar potentials~\cite{Dzhunushaliev:2008bq,Jaramillo:2023pny,Carvente:2019gkd}. In contrast, comparatively little attention has been given to the role of ghost scalar fields within realistic compact objects that are otherwise supported by ordinary matter. In particular, it is not yet well understood whether a ghost sector can coexist with standard matter in a stable equilibrium configuration, or whether its inclusion generically triggers dynamical instabilities that compromise the viability of the system.

In a previous work~\cite{Jaramillo:2023twi}, configurations were constructed in which ghost matter is gravitationally confined within an otherwise canonical boson star. By localizing the ghost sector to a finite region, its properties could be examined in a controlled framework, and regular, non-divergent solutions were obtained at the classical level. The ghost component was modeled as a real scalar field $\chi$ whose kinetic term in the Lagrangian carries the opposite sign relative to that of a canonical scalar field. The potential adopted in that study included both a mass term and a quartic self-interaction, namely $V(\chi)=-\mu^2\chi^2 + \frac{\lambda}{2}\,\chi^4$. In the limit where the exterior canonical boson star is absent and the configuration reduces to a purely central ghost soliton, both contributions to the potential are required in order to obtain static solutions~\cite{Dzhunushaliev:2008bq}.

Building on this framework, we now extend the analysis to a more realistic and astrophysically relevant setting: neutron stars composed of ordinary nuclear matter. Unlike boson stars, neutron stars provide a well-established strong-field laboratory in which the interplay between standard matter and exotic sectors can be systematically investigated. Using the techniques developed in the study of exotic compact objects~\cite{bezares2025exotic} and mixed stars, namely neutron stars with bosonic cores~\cite{valdez2013dynamical,brito2016interaction,bezares2019gravitational,DiGiovanni:2020frc,di2021stabilization,lazarte2025gravitational}, we investigate equilibrium configurations in which the ghost sector, modeled here as a complex scalar field with a kinetic term of opposite sign in the Lagrangian density, occupies the central region of a neutron star. In this sense, the ghost component plays a role analogous to the bosonic core in standard mixed-star models. For the scalar potential, we adopt a massive term together with a quartic self-interaction, characterized by the scalar field mass $\mu$ and coupling constant $\lambda$, with $\mu$ entering with the opposite sign compared to the case of a standard (mini) boson star.

Our numerical results demonstrate that ghost scalar matter can indeed be gravitationally trapped within a neutron star. This allows us to conclude more generally that exotic matter can be confined in gravitational potential wells, despite violating standard energy conditions. Beyond the construction of equilibrium solutions, we also perform fully dynamical evolutions of these configurations, resulting in nontrivial behavior that will be detailed below.

Our analysis further suggests that dark-energy-like components, even when modeled as fields that violate the weak or null energy conditions~\cite{Jaramillo:2023twi}, could in principle be gravitationally confined, provided that more extreme violations (such as those exhibited by ghost matter) can be consistently accommodated in compact objects. In this sense, our configurations share qualitative similarities with gravastar models~\cite{mazur2004gravitational}, providing a dynamically accessible and fully relativistic framework that captures several of their key features while remaining amenable to detailed numerical analysis.

This work is organized as follows. In Sec.~\ref{stat}, we construct equilibrium configurations consisting of an Oppenheimer–Volkoff neutron star coupled to a complex scalar field with an overall opposite sign in the Lagrangian density, representing the ghost sector. The two components interact exclusively through gravity: both contribute to the spacetime geometry, and the resulting geometry consistently determines their respective field equations. We then solve the coupled system using two independent numerical codes~\cite{DiGiovanni:2020frc,lazarte2025gravitational}, constructing families of solutions parameterized by the central amplitude of the ghost field and assessing its impact on the global properties of the configuration. In Sec.~\ref{sec:evolution}, we employ a well-tested one-dimensional numerical relativity code NADA1D~\cite{Montero:2012yr,Sanchis-Gual:2014ewa,DiGiovanni:2020frc,lazarte2025gravitational} to evolve selected configurations by solving the full Einstein–(ghost, complex) Klein–Gordon–Euler system. These evolutions show that several solutions are dynamically stable and exhibit distinctive oscillatory features, indicating that ghost matter can remain confined within the neutron star’s gravitational potential and potentially lead to observable consequences.
Finally, in Sec.~\ref{sec:discussion}, we summarize and discuss our results. Throughout this work, we adopt Planck units.

\section{Basic equations for the equilibrium configurations} \label{stat}

The formalism for the construction of  equilibrium configurations of fermion-exotic boson stars relies on the choice of a spherically symmetric metric in Schwarzschild coordinates
\begin{equation} \label{Sph-Sta_metric}
ds^2 = -\alpha(r)^2 dt^2 + \tilde{a}(r)^2 dr^2 + r^2 ( d\theta^2 + \sin{\theta}^2 d\varphi^2),
\end{equation}
written in terms of two geometrical functions $\tilde{a}(r)$ and $\alpha(r)$. 

\subsection{Field equations}

We analyze configurations consisting of two types of matter: a fluid component and an exotic component described by a complex scalar field $\chi$. Accordingly, Einstein’s equations, $G_{\mu\nu} = 8\pi T_{\mu\nu}$, involve a total stress–energy tensor of the form
$T_{\mu\nu}=T^{\rm fluid}_{\mu\nu} + T^{\rm SF}_{\mu\nu}$,
where $T^{\rm fluid}_{\mu\nu}$ denotes the stress–energy tensor of the fluid, which models the neutron star:
\begin{equation}
T^{\rm fluid}_{\mu\nu}=\left(\rho\,\left(1+\epsilon\right) +P\right)\,u_\mu\,u_\nu + P\,g_{\mu\nu},  \label{eq:Tmunu_f}
\end{equation}
with $\rho$ the rest mass density of the fluid, $\epsilon$ its internal energy and $P$ its pressure, and  $T^{\rm SF}_{\mu\nu}$ denotes the stress energy tensor of the ghost matter, described by a complex scalar field $\chi$:
\begin{equation}
 T^{\rm SF}_{\mu\nu}=-\frac{1}{2}\,\left[\nabla_\mu\,\chi\,\nabla_\nu\,\chi^* + \nabla_\mu\,\chi^*\,\nabla_\nu\,\chi - g_{\mu\nu}\,\left(g^{\alpha\beta}\,\nabla_\alpha\,\chi\,\nabla_\beta\,\chi^* - V(|\chi|^2)\right)\right]. \label{eq:Tmunu_g} 
\end{equation}
For the exotic matter we consider an harmonic time dependence, $\chi(t, r) = \chi(r) e^{-i\omega\, t}$, where $\omega$ is its eigenfrequency. Notice that making such frequency equals to zero, makes the scalar field real, so that the real cases are also included in our description. Regarding the scalar potential, we consider a massive scalar field and include the quartic self-interaction potential:
\begin{equation}
    V(|\chi|^2)=-\mu^2|\chi|^2+\frac{\lambda}{2}|\chi|^4,
\end{equation}

Assuming a static fluid, $u^{\beta}=(-1/\alpha,0,0,0)$, Einstein's equations with the double matter source, lead to the following ordinary differential equations (ODEs). We are following the procedure presented in  \cite{DiGiovanni:2020frc, Bustillo:2020syj, Jaramillo:2023twi}.

\begin{eqnarray}
\frac{d\tilde{a}}{dr} & = &\frac{\tilde{a}}{2}\left(\frac{1-\tilde{a}^{2}}{r} + 4\pi r \biggl[2\tilde{a}^{2}\rho(1+\epsilon) - \biggl(\Sigma^2 + \biggl(\frac{{\omega}^{2}}{\alpha^{2}}+{\mu}^{2}-\frac{\lambda}{2}|\chi|^{2}\biggl) \tilde{a}^{2}|\chi|^{2}  
 \biggl) \biggl]\right.\biggl), \label{s1}
  \\
\frac{d\alpha}{dr} & = &\frac{\alpha}{2}\left(\frac{\tilde{a}^{2}-1}{r} + 4\pi r \biggl[2\tilde{a}^{2}\,P - \biggl(\Sigma^2 + \biggl(\frac{{\omega}^{2}}{\alpha^{2}} - {\mu}^{2} + \frac{\lambda}{2}|\chi|^{2}\biggl) \tilde{a}^{2}|\chi|^{2}
 \biggl) \biggl] \right.\biggl), \label{s2}
\end{eqnarray}
and for the Klein Gordon equation we have considered two first order equations:
\begin{eqnarray}
\frac{d\chi}{dr} &=& \Sigma , \label{s3}
\\
\frac{d\Sigma}{dr} & =& -\left(1+\tilde{a}^{2} + 4\pi r^{2}\tilde{a}^{2}\,\biggl(P - \rho(1+\epsilon) + \biggl({\mu}^{2}\,|\chi|^{2}-\frac{\lambda}{2}\,|\chi|^{4} \biggl) \right)\biggl)\frac{\Sigma}{r} - \left(\frac{{\omega}^{2}}{\alpha^{2}} - {\mu}^{2} + \lambda\,|\chi|^{2}\right)\tilde{a}^{2}\chi, \label{s4}
\end{eqnarray}
where we have made use of the Einstein's equations to express the derivatives of the metric components in terms of the matter ones. For the Euler equation we have
\begin{equation}
\frac{dP}{dr} = -[\rho(1+\epsilon)+P]\frac{d\ln\alpha}{dr},
\label{s5}
\end{equation}
and the system is closed by the equation of state:
\begin{equation}
P=K\,\rho^\Gamma=\left(\Gamma - 1\right)\,\rho\,\epsilon \, . \label{eq:EoS}  
\end{equation}

We consider the following boundary conditions of asymptotic flatness, $a|_\infty=\alpha|_\infty=1, \chi|_\infty=0$, and demand regularity at the origin: $a(0)=1, P(0)=K\rho_0^\Gamma,  \rho(0)=\rho_0, \chi(0)= \chi_0$.

We will consider  the particular choice of $\Gamma=2$  and $K=100$, which corresponds to masses and compactness in the range of neutron stars. Also, 
we take $\mu=1$ and explore several values of the self interaction constant, $\lambda$. The system is set to be solved.

We recall that from the Arnowitt-Deser-Misner ADM mass 
\begin{equation}
M_{T} = \lim_{r \to \infty} \frac{r}{2}  \left( 1 - \frac{1}{\tilde{a}(r)^2} \right),
\end{equation} 
several functions can be obtained which are useful to characterize the self gravitational body. Notice that we have two characteristic radius: the radius $R_{99}^{\chi}$, defined as the value of the radius that enclosed $99\%$ of the absolute value of energy density of the ghost field; and the radius where the rest mass density of the fluid becomes $0$, denoted by $R_{N}$. The compactness of the mixed star is defined as $M_T/\max\{R_{99}^{\chi},R_{N}\}.$ Also we will have several densities: the total density $\rho_T$, being the integral over the volume of the ADM mass, the fluid density, $\rho$, obtained from the stress energy tensor, Eq.~\eqref{eq:Tmunu_f}, and the density directly related to the ghost field, from the corresponding stress energy tensor, Eq.~\eqref{eq:Tmunu_g}, we obtain:
\begin{equation}
 \rho_\chi=-\frac12\,\left(\frac{{\Sigma}^2}{a^2} + \left(\frac{\omega^2}{\alpha^2} + \mu^2 - \frac{\lambda}{2}\chi^2 \right)\,\chi^2\right),   
\end{equation} 
with $\chi=\chi(r)$. In describing the mixed star, we will use these definitions. 
%\clearpage

\section{Initial configurations}
\label{Sec:Initial_c}
To solve the system of equations Eqs.~(\ref{s1}-\ref{eq:EoS}), with the appropriate boundary conditions and value $\Gamma, K$ and $\mu$ described above, we have used two independent numerical codes. In both, a value for the $\lambda$ parameter is chosen, and the system of equations with the boundary conditions become an eigenvalue problem for $\omega$. We use the shooting method to find the solutions. To construct the configurations, we either fix a values for the central density of the fluid, $\rho_0$, and then explore for a range of values for the the scalar field at the center of the configuration, $\chi_0$, or the other way around, fix $\chi_0$ and explore $\rho_0$. The code determines the scalar field and density profiles, the fluid pressure and the metric coefficients.

As a first step, we generate several configurations of neutron stars  for several values of the central density  $\rho_0$ (A typical value is $\rho_0=1.28\,\times\,10^{-3}$), and explore a range of values of the magnitude of the ghost scalar field at the center, $\chi_0$, compute the corresponding mixed configuration, and determine its properties: total mass of $M_T$, the radii  $R_N$ and  $R_{99}^{\chi}$ defined above, and compactness. The maximum value of $\chi_0$ of the mixed star corresponds to $\omega=0$, and as $\omega$ increases, $\chi_0$ decreases, $\chi_0=0$ corresponds to the neutron star with $\rho_0$.
%we also generate configurations fixing the value of the ghost field at the center, $\chi_0$ and sweeping the value of the fermiomic matter there. 

\begin{figure}[H]
	\begin{centering}
		\includegraphics[scale=.23]{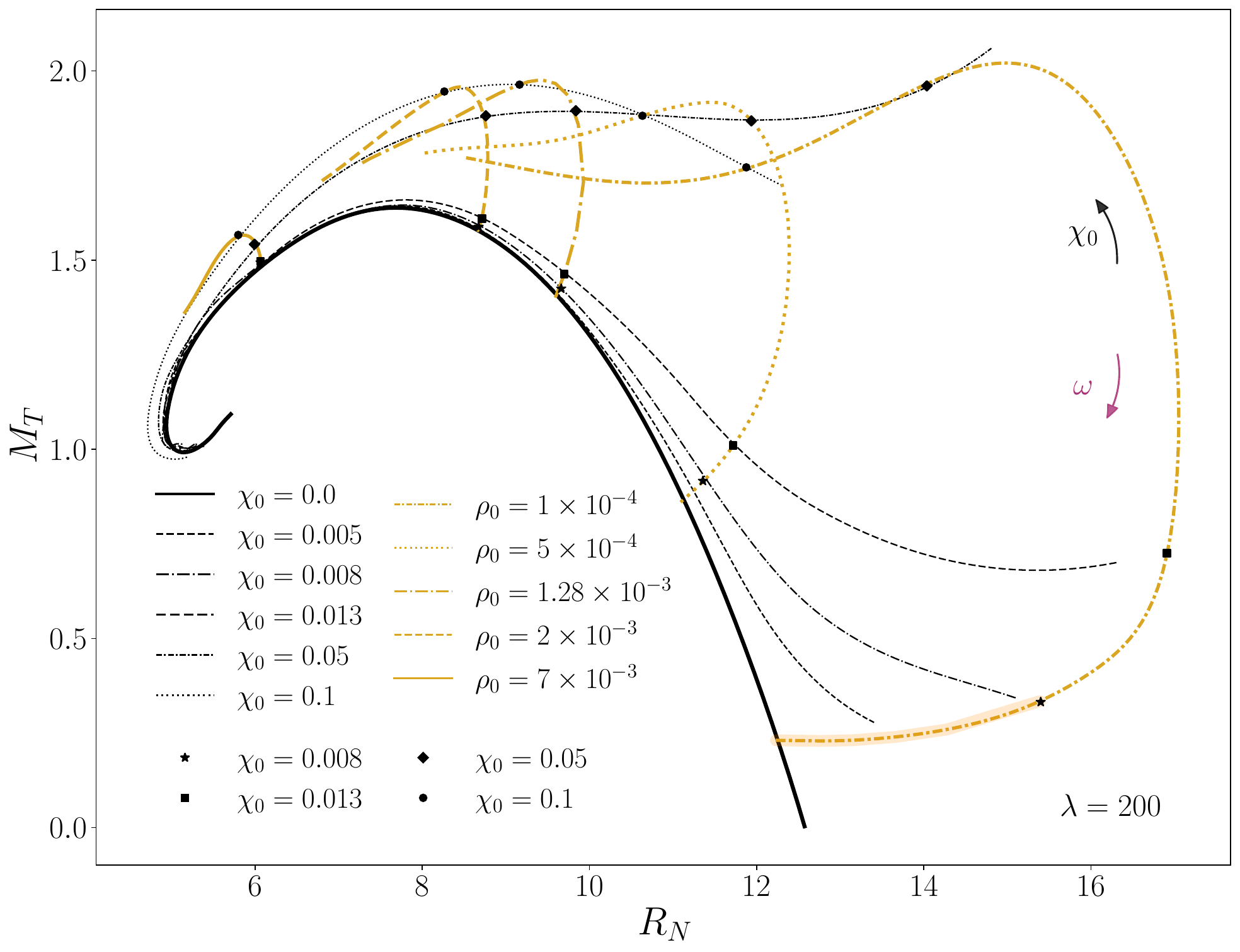} \hspace{0.1cm}
        \includegraphics[scale=.23]{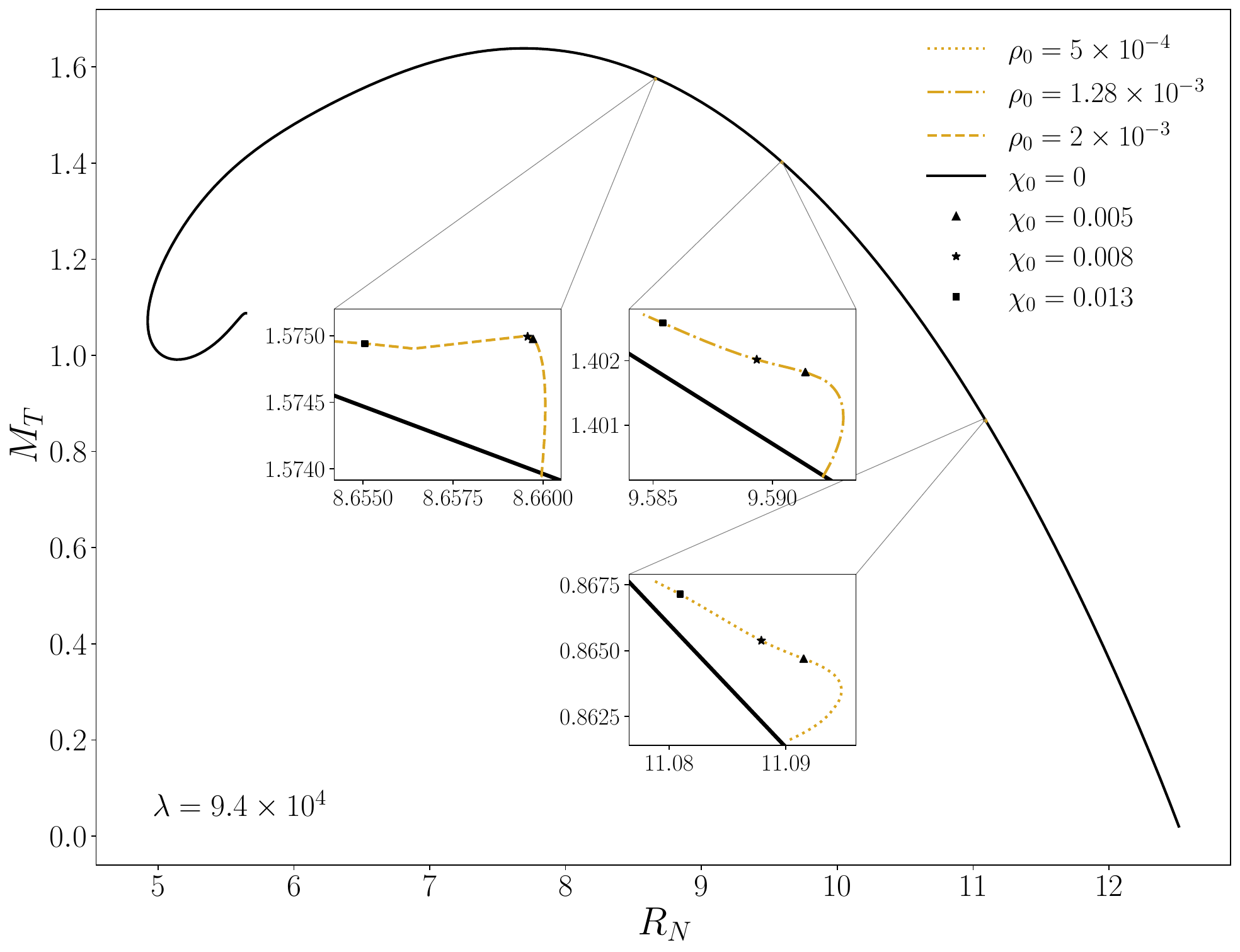}
		\par
	\end{centering}
	\caption{Neutron star configurations coexisting with exotic matter. We present 
the total mass $M_T$ of the configuration as a function of the neutron star radius $R_N$, for two values of the self interaction parameter $\lambda$. 
The solid line corresponds to neutron stars without exotic nuclei. 
Gold curves denote families of mixed stars with constant $\rho_0$ and varying $\chi_0$, 
while dotted black curves denote constant $\chi_0$ and varying $\rho_0$. 
In most constructed configurations, the ghost matter is confined within the fluid 
($R_{99}^N < R_N$); however, for $\lambda = 200$, configurations in which the 
opposite occurs are also found and are highlighted in yellow. In the large self-interaction regime, configurations belonging to a family with fixed $\rho_0$ show that both $M_T$ and $R_N$ depend on $\chi_0$; nevertheless, their values are nearly indistinguishable from those of the corresponding pure neutron star with the same $\rho_0$.} 
\label{fig:MvR02}
\end{figure}

\begin{figure}[H]
	\begin{centering}
		\includegraphics[scale=.35]{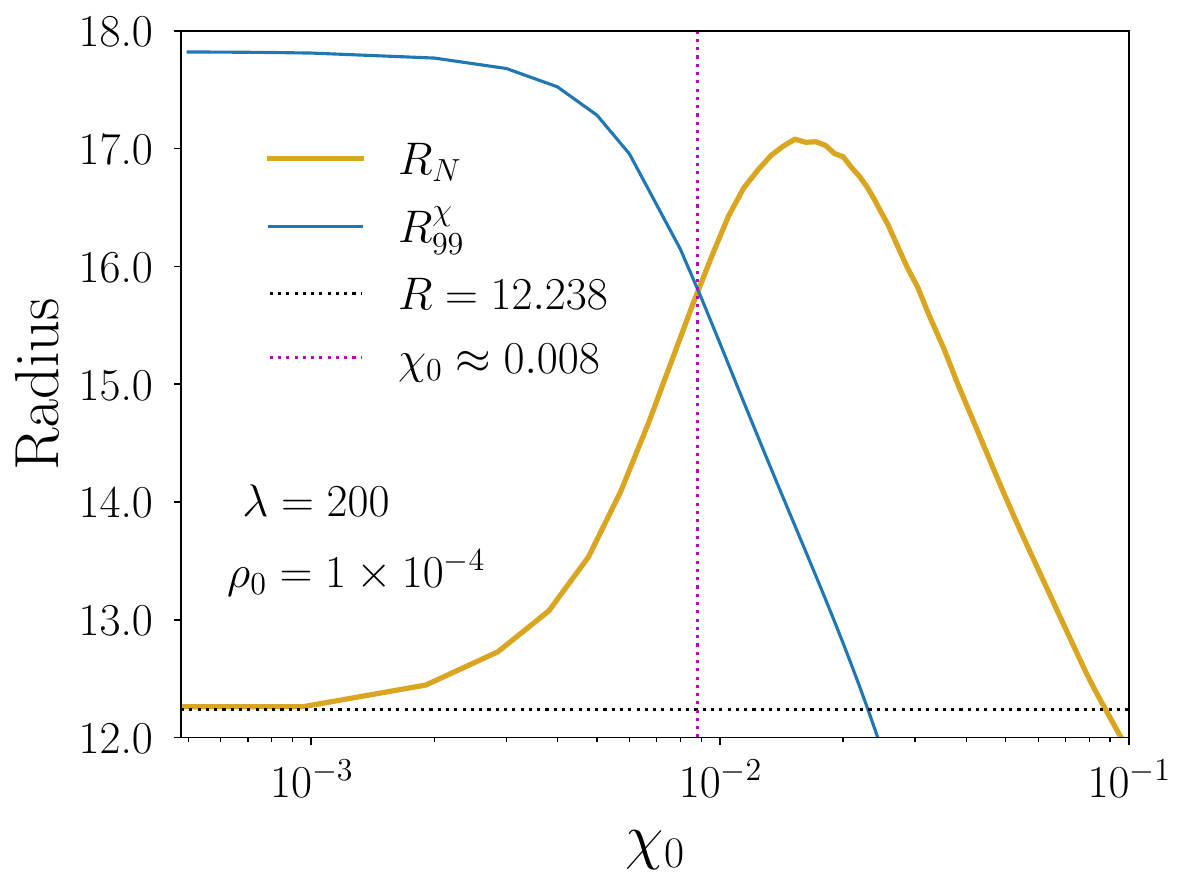} \hspace{0.1cm}
        \includegraphics[scale=.35]{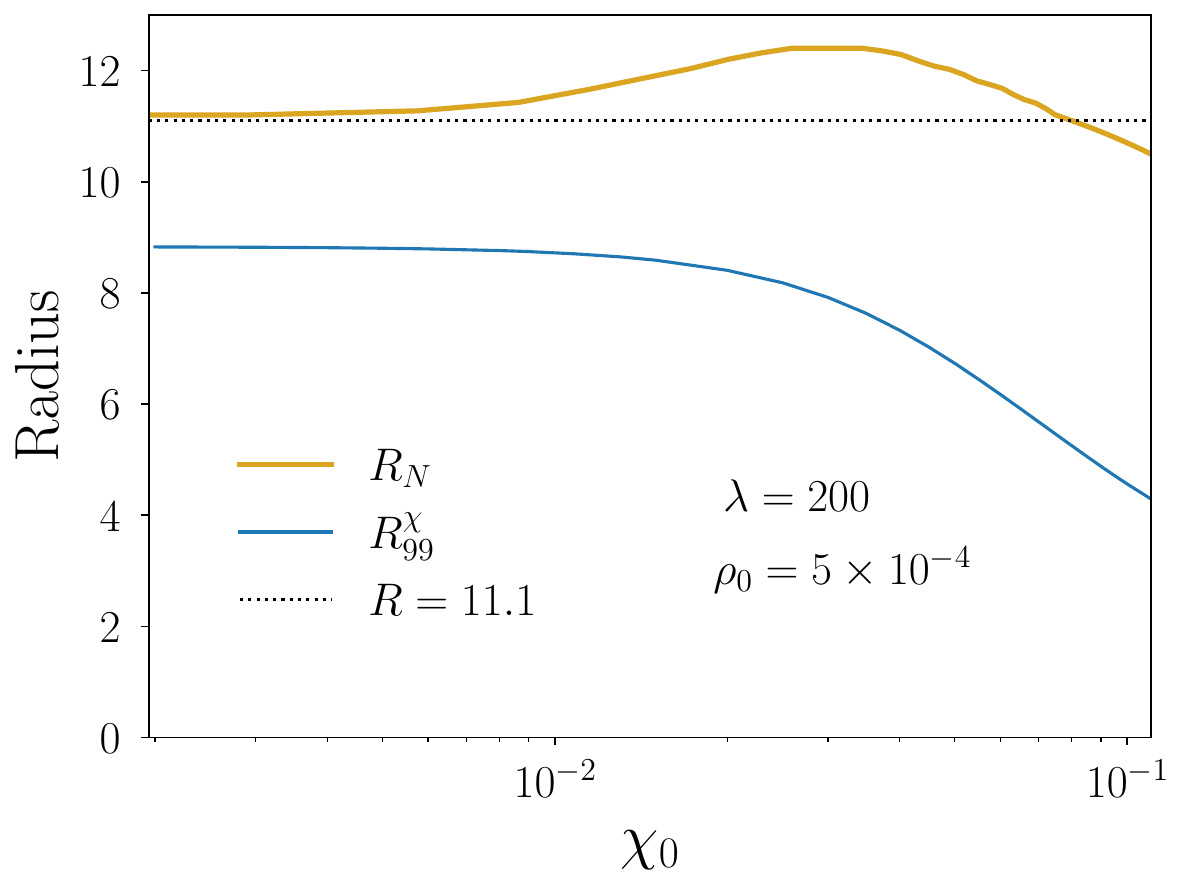}
		\par
	\end{centering}
	\caption{Neutron star radius, $R_N$ and phantom nuclei one, $R^{\chi}_{99}$, (see definitions in the text) as functions of value of the phantom field at the center, $\chi_0$. We present two families of mixed configurations with $\lambda = 200$ and fixed $\rho_0$: left panel, 
$\rho_0 = 1 \times 10^{-4}$; right panel, $\rho_0 = 5 \times 10^{-4}$. 
The black dotted line indicates the radius of the corresponding neutron star without exotic nuclei
for each value of $\rho_0$. In the case $\rho_0 = 1 \times 10^{-4}$, 
configurations exist in which the fluid component is trapped by the phantom field.  
} 
\label{fig:RNvRchi}
\end{figure}

In Fig.~\eqref{fig:MvR02} we present the behavior of the total mass $M_T$, against the radius of the fluid component, $R_{N}$. We do so for two values of the self interaction constant, $\lambda=200$, left panel, and  $\lambda=9.4\,\times\,10^4$, right panel.  In both panels, the solid line describes neutron stars without exotic nuclei. For the case of $\lambda=200$, it is remarkable that there exist mixed configurations whose profiles differ substantially from those of the neutron stars from which they were originally constructed. There are mixed configurations for which larger values of the total mass of the configurations correspond to larger $R_{N}$. There are cases where the total mass $M_T$ reaches a value close to $2.0$, significantly larger than the maximal value of a neutron star without phantom nuclei; such behavior is also shown for fixed central values of the fermionic density, $\rho_0$; as such value increases, there are configurations which reach the above mentioned large value of the maximal mass, but for smaller  radius of the neutron star, $R_{N}$. In general, for $\lambda=200$, as the value of $\chi_0$ increases, the total mass value, $M_T$, increases, reaches a maximum, and then decreases, this is also shown in Fig.~\eqref{fig:RNvRchi}. For mixed configurations generated from not so compact neutron stars, $R_N$ increases as $\chi_0$ does, reaches a maximum, and then decreases, while for more compact initial neutron stars, as $\chi_0$ increases, $R_N$ decreases. 

In Fig.~\eqref{fig:MvR02}, we indicate with dots the points where the configurations for fixed $\chi_0$ and sweeping $\rho_0$ intersect with the configurations obtained by fixing $\rho_0$ and sweeping $\chi_0$. In general, in the mixed configurations studied, 
the ghost scalar field is confined within the fluid $R_{99}^{\chi} < R_N$, however, for $\lambda=200$ it is noteworthy that we have found  configurations in which the opposite occurs $R_N< R_{99}^{\chi}$, the fluid forming a core enclosed by a ghost-field envelope; these configurations are highlighted in yellow in Fig.~\eqref{fig:MvR02}. In Fig.~\eqref{fig:RNvRchi} both radius,  $R_N$ and $R_{99}^{\chi}$ are shown as functions of $\chi_0$ for $\lambda=200$ and for two values of the central fermionic density, $\rho_0=1\times 10^{-4}$ and $5\times10^{-4}$. 

In Fig.~\eqref{fig:MvR02}, it can also be observed that the differences in $M_T$ and $R_N$ among configurations belonging to the same family (with fixed $\rho_0$) decrease as $\lambda$ increases; in this regime, the family of configurations is almost not noticeable affected. In the case of $\lambda=9.4\times 10^4$,  for mixed configurations, with $\rho_0$ constant, the total mass, $M_T$ and $R_N$ are also functions of $\chi_0$, however, for a given $\rho_0$, the maximum attainable value of $\chi_0$ (that for which $\omega=0$), is smaller than that corresponding to the case $\lambda = 200$.

In Fig.~\eqref{fig:RNvRchi} we present the radii as a function of the central value of the ghost field, $\chi_0$ and for $\lambda=200$, for two values of the fluid central value. For $\rho_0=1\,\times\,10^{-4}$, left panel, we see that for small values of $\chi_0$, the ghost radius is larger than the fluid one, the neutron star is enclosed by the ghost field. However, the ghost radius diminishes as $\chi_0$ increases, and the fluid radius first increases and then decreases, so that for $\chi_0>0.009$ the ghost nuclei is inside the neutron star and diminishes very fast. For $\rho_0=5\,\times\,10^{-4}$, right panel, all the mixed configurations obtained in our numerical analysis satisfy $R_{99}^{\chi} < R_N$, suggesting that, within the explored parameter space, the exotic component is fully enclosed by the neutron star in this case.

\begin{figure}[H]
	\begin{centering}
		\includegraphics[scale=.25]{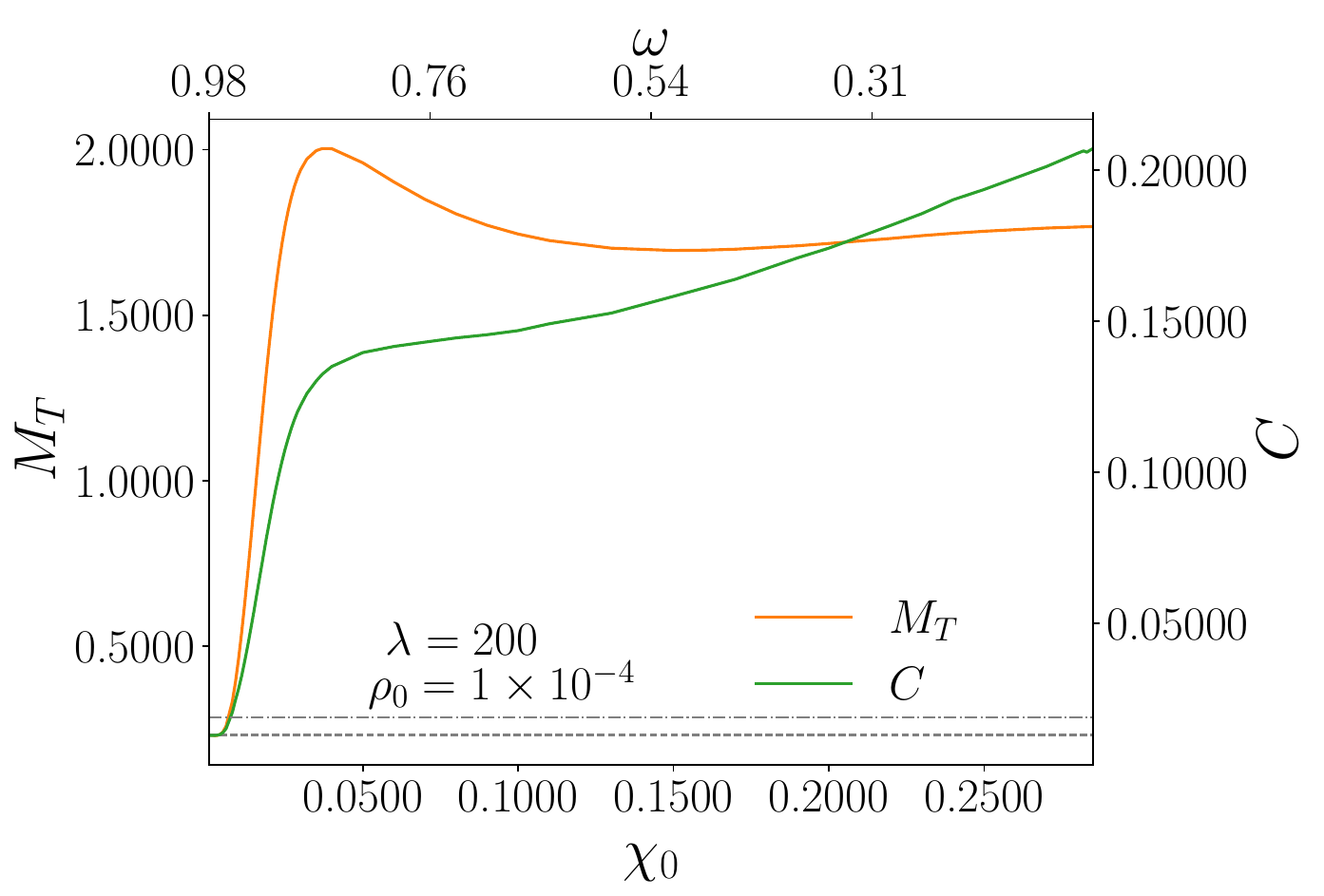} 
        \includegraphics[scale=.25]{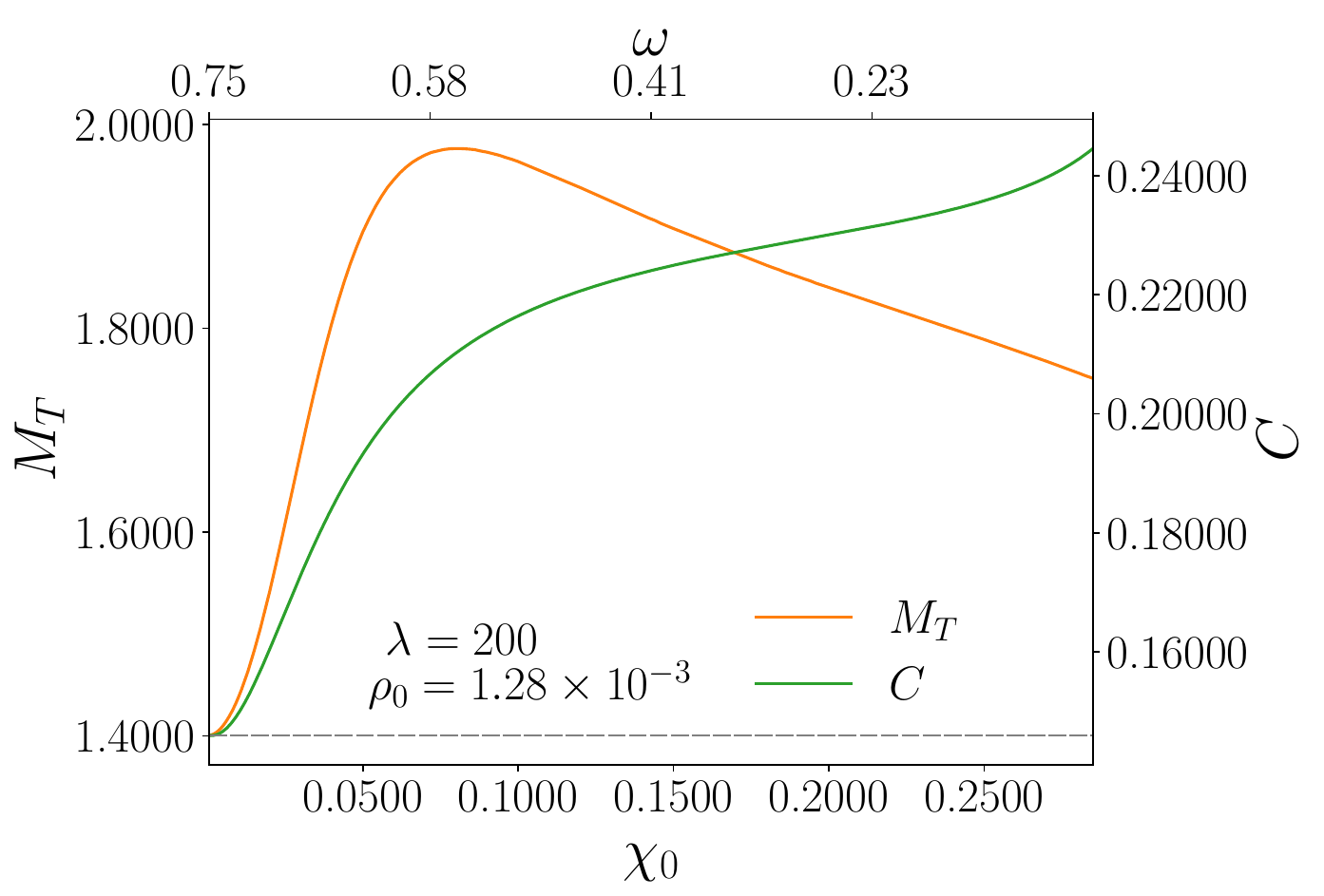}
        \includegraphics[scale=.25]{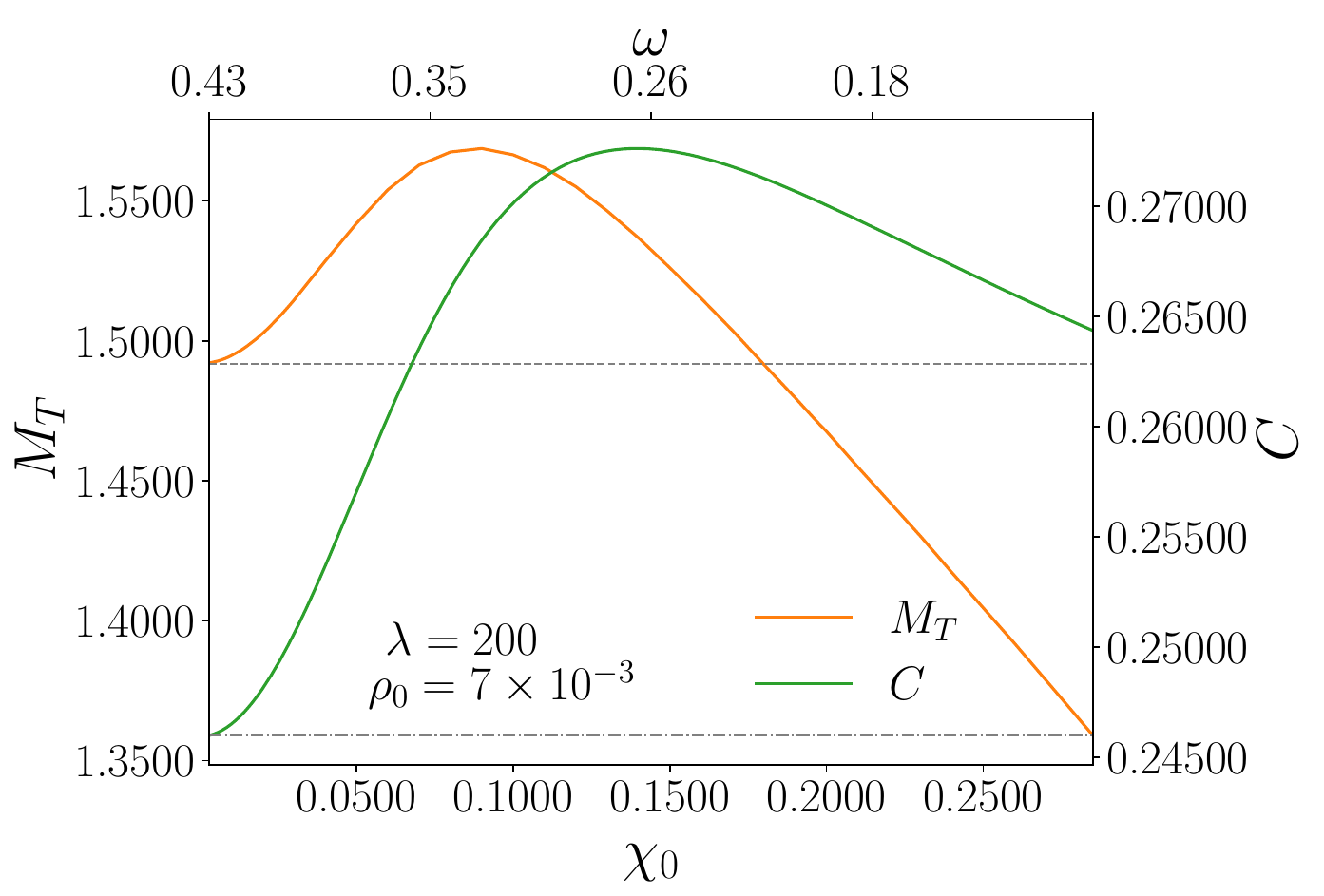}\hspace{0.5cm}
		\par
        \includegraphics[scale=.25]{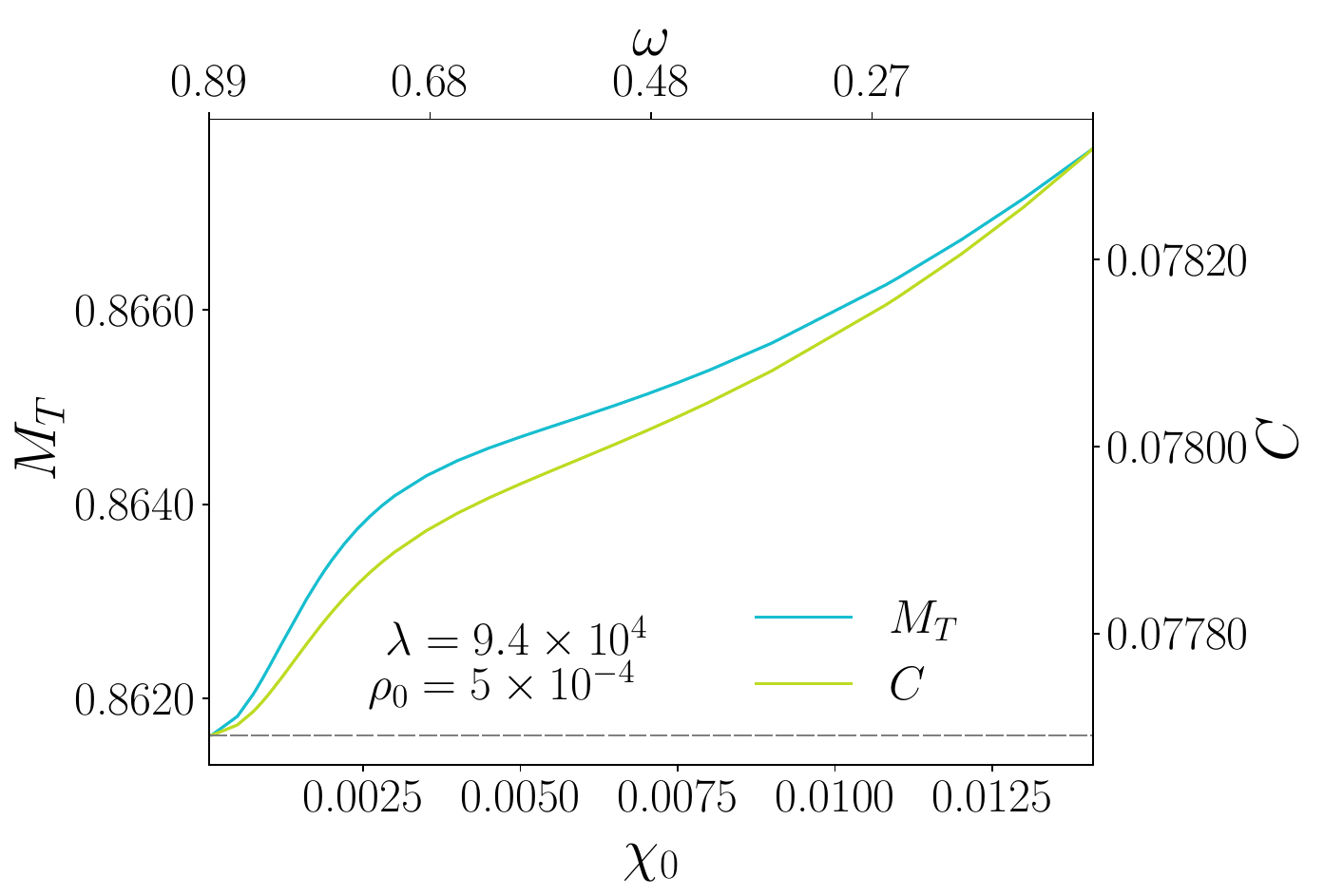} 
        \includegraphics[scale=.25]{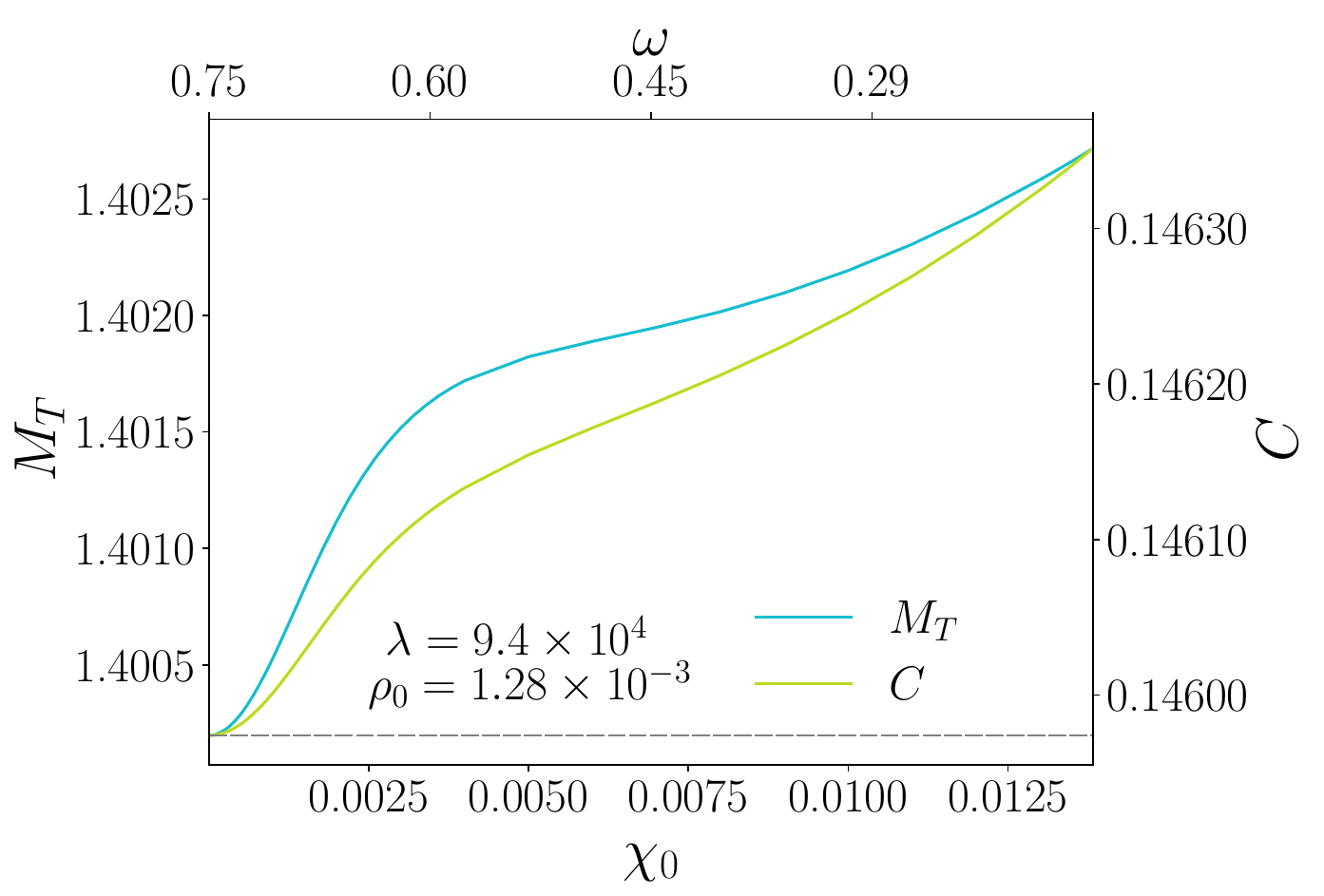}
        \includegraphics[scale=.25]{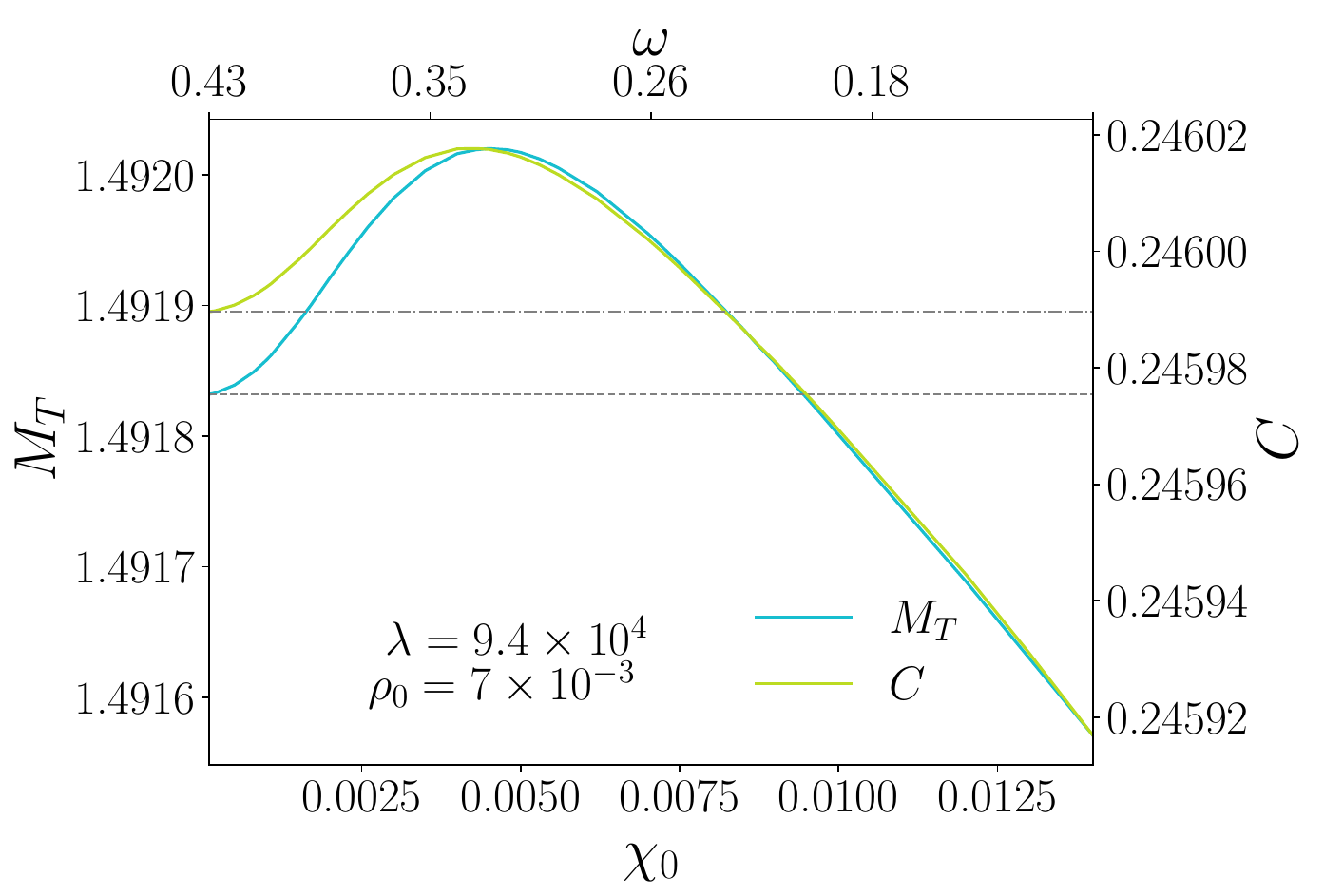}
        \par
	\end{centering}
	\caption{Total mass and compactness as functions of $\chi_0$ and $\omega$ 
for mixed configurations with fixed $\lambda$ and $\rho_0$. 
Dashed lines denote the reference values at $\chi_0 = 0$. 
For small $\rho_0$, lower $\lambda$ yields significantly larger masses 
and compactness than the pure neutron star, whereas for 
$\lambda = 9.4 \times 10^4$ the increase is modest. 
In families with $\rho_0=7\times 10^{-3}$ for large enough values of $\chi_0$, there exist configurations for which the total mass, $M_T$, becomes smaller 
than that of the pure neutron star. }
\label{fig:Mvsomega}
\end{figure}

In Fig.~\eqref{fig:Mvsomega} we present profiles of the total mass and the compactness, against $\chi_0$ and $\omega$, for families of mixed configurations with fixed values of $\rho_0$ and $\lambda$. 
As the $\chi_0$ value increases, the configuration shifts towards smaller values of $\omega$, also reducing the possible range of frequencies.
For reference, the mass and the compactness of the corresponding pure neutron stars are indicated. For the smaller value, $\lambda=200$, the changes for a given central value $\chi_0$ are clearly noticeable with respect to the case without exotic nuclei. The changes in the profiles for large value of $\lambda=9.4\,\times\,10^4$, are not so noticeable, producing changes in the third decimal cipher. For all families, $M_T$ starts to grow as soon as $\chi_0$ is switched on, the same is true for the compactness except for such configurations for which $R_N <R_\chi$, see top left panel of Fig.~\eqref{fig:Mvsomega}; in that case, the mixed configurations are less compact than the corresponding pure neutron star. For small values of $\rho_0$, mixed configurations within a family characterized by fixed $\rho_0$ always exhibit a total mass larger than that of the corresponding pure neutron star. Nevertheless, for families with $\rho_0 = 7 \times 10^{-3}$, for the two values of $\lambda$, and sufficiently large $\chi_0$, configurations can be found for which the total mass $M_T$ becomes smaller than that of the pure neutron star.
Notice, however, that the value for the configuration with maximal mass is almost not affected by the central value of the ghost field, and the maximum value of the compactness does not changes appreciably either.
\begin{figure}[H]
	\begin{centering}
		\includegraphics[scale=.4]{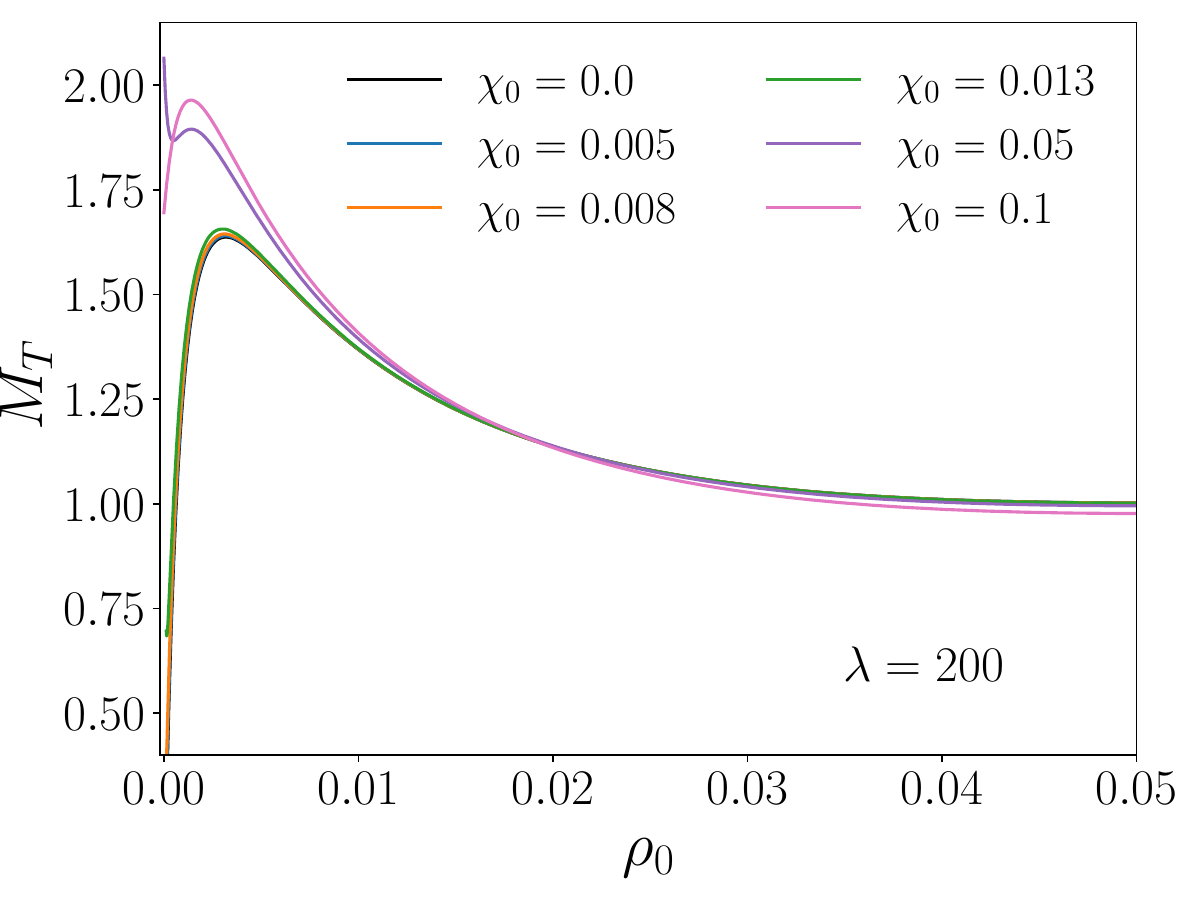} 
		\par
	\end{centering}
	\caption{Families of neutron stars configurations coexisting with exotic matter.  We present behavior of the total mass $M_T$ of the star vs the central density $\rho_0$ for $\lambda=200$ and fixed values of the phantom scalar field at the center.}
\label{fig:M_Tvscentraldensity}
\end{figure}

In Fig.~\eqref{fig:M_Tvscentraldensity} we present families of neutron stars configurations coexisting with exotic matter with $\lambda=200$ constructed by fixing a value of $\chi_0$ and solve for a range of values of $\rho_0$. As expected, the change in the total mass profile is clearly noticeable when the value of the fermionic density at the center is smaller that the corresponding value of the ghost matter.

Overall, we see that there are indeed families of configurations of neutron stars coexisting with exotic matter. The configurations do not need any fine tuning. We have explored configurations within the range of $\chi_0 \in [0.005, 0.1]$ for the value of the ghost field at the center. 

\begin{figure}[H]
	\begin{centering}
        \includegraphics[scale=.28]{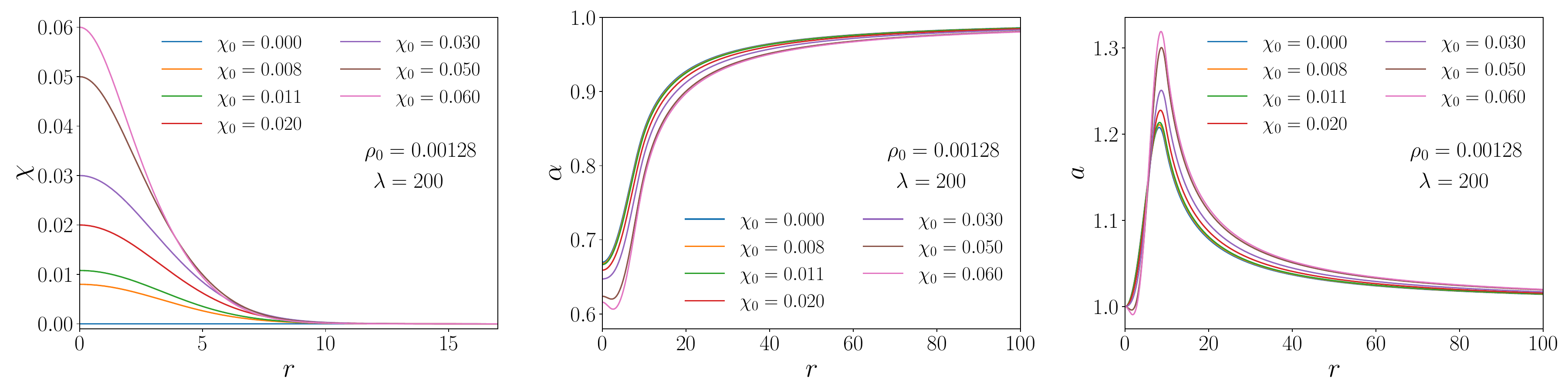} \hspace{0.1 cm}
    \includegraphics[scale=.28]{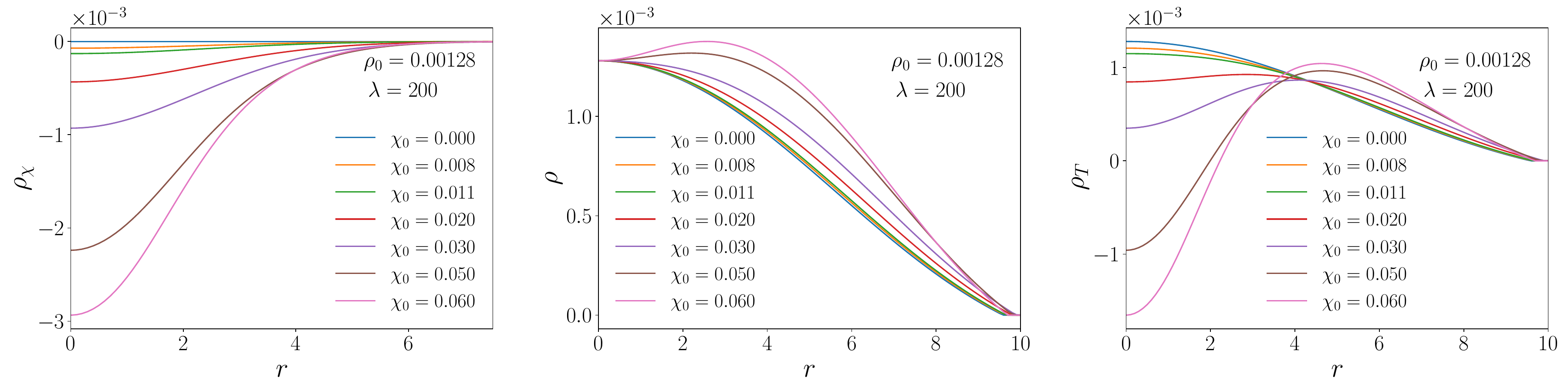} 
  \par
	\end{centering}
	\caption{Profiles of $\chi(r)$, and the metric coefficients  $\alpha(r), a (r)$, for a configuration with $\rho_0=1.28\times 10^{-3}, \lambda=200$, and several values of the phantom field at the center. The fermionic density, $\rho(r)$, the phantom scalar field density $\rho_{\chi}$ and $\rho_T$ are also shown for the same configurations}
\label{fig:e-NS-L200 A}
\end{figure}

We proceed to the analysis of specific configurations, in order to enhance the physical properties of the mixed  neutron stars / phantom field. 
In Fig.~\eqref{fig:e-NS-L200 A} we present profiles of the phantom field, $\chi(r)$, and the metric coefficients  $\alpha(r), a (r)$, for a configuration with $\rho_0=1.28\times 10^{-3}, \lambda=200$, and several values of the phantom field at the center. The fermionic density, $\rho(r)$, the phantom scalar field density $\rho_{\chi}$ and the total density, $\rho_T$ are also shown for the same configurations. 

\begin{figure}[H]
	\begin{centering}
		\includegraphics[scale=.28]{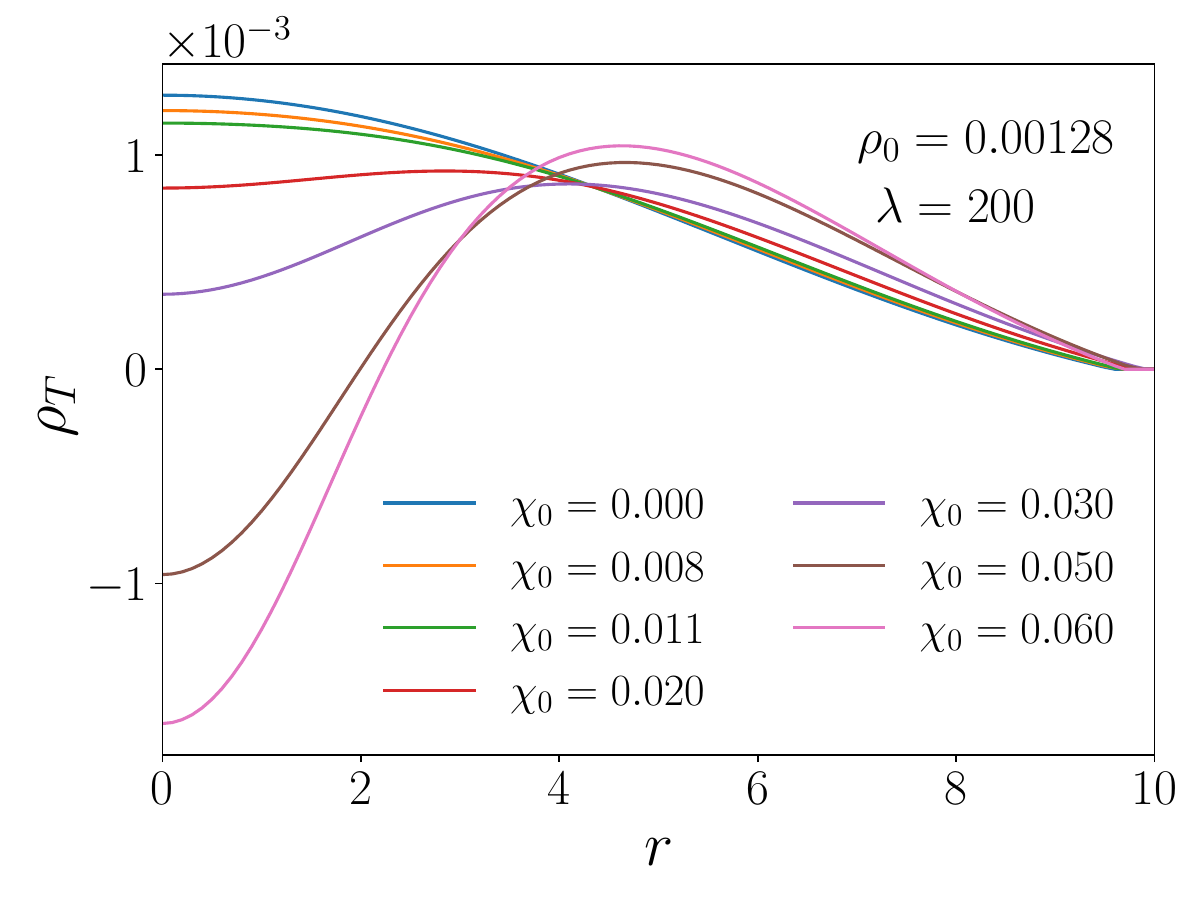} \hspace{0.1cm}
        \includegraphics[scale=.28]{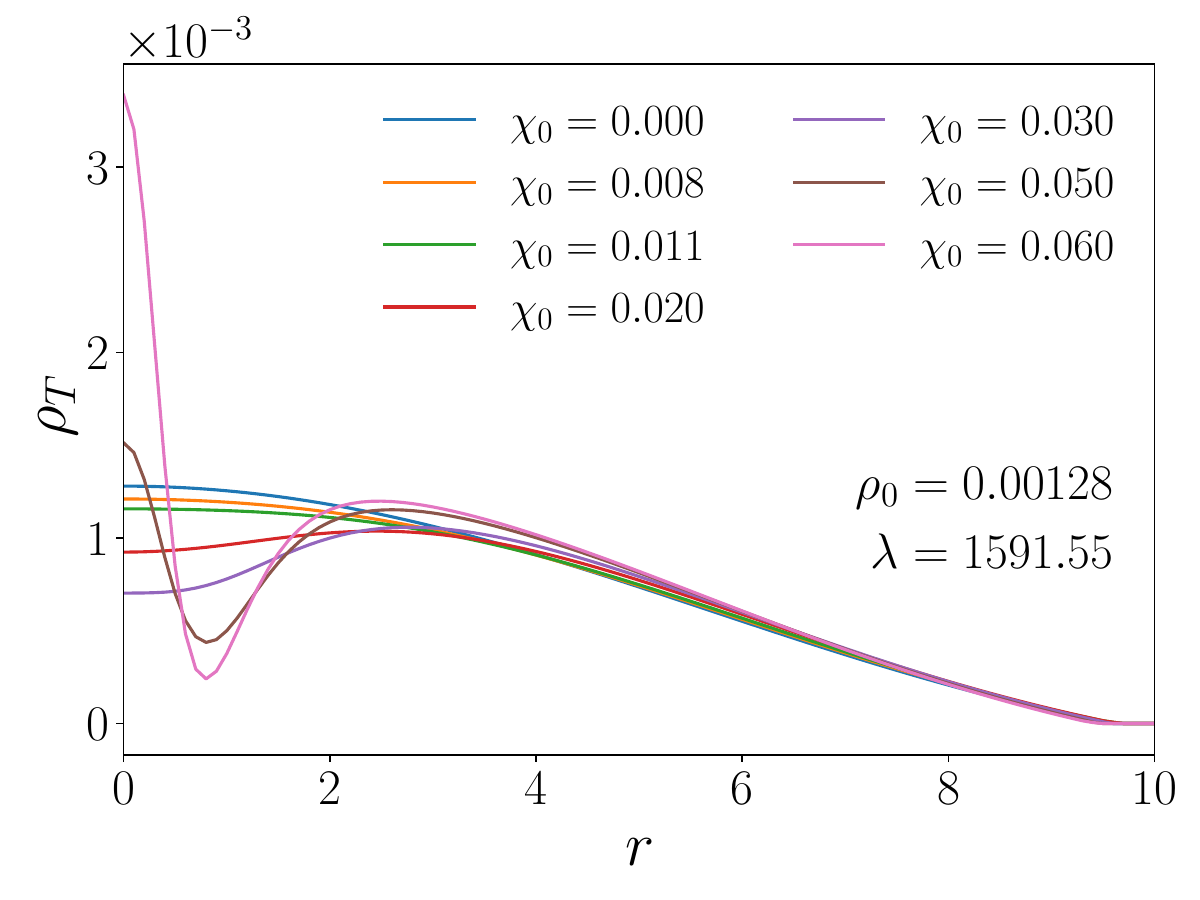}
        \hspace{0.1cm}
        \includegraphics[scale=.28]{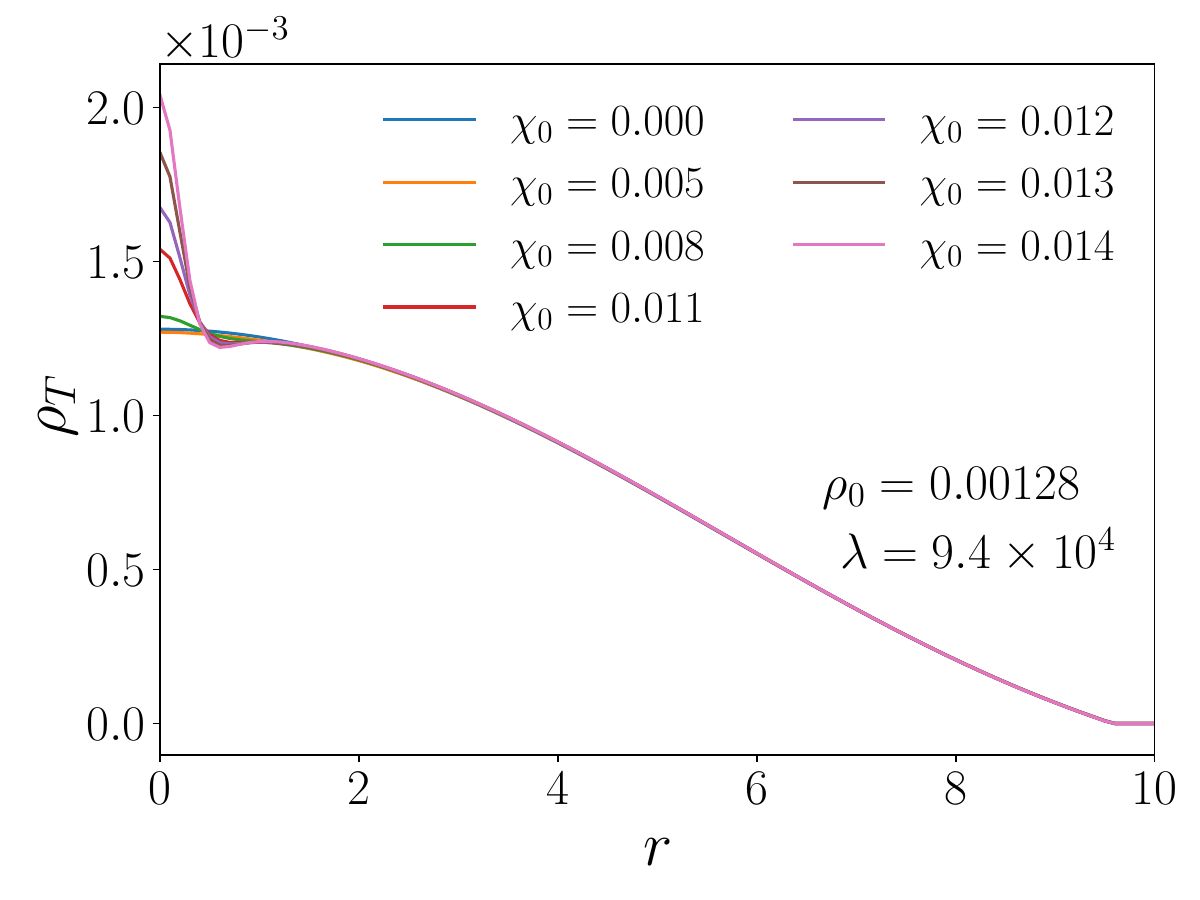}
        
		\par
	\end{centering}
	\caption{We present the profile for the total density of the configurations as a function of the radius. We do so for a fixed values of the central fermionic density, $\rho_0=0.00128$ and explore the cases for several values of the phantom scalar field at the center. We present the profile for the case when the self interaction parameter $\lambda=200$, left, $\lambda=1591.55$, center, and $\lambda=9.4\,\times\,10^4$, right.}
\label{fig:rhos}
\end{figure}

In Fig.~\eqref{fig:rhos}, we present the total density profile, $\rho_T$, for selected configurations corresponding to the fiducial central neutron-star density $\rho_0 = 1.28 \times 10^{-3}$. For comparison, the density profile of the pure neutron star (without ghost nuclei) is also shown. In the case $\lambda = 200$ (top-left panel), the ghost field shifts the maximum of $\rho_T$ away from $r = 0$—the location of the maximum for the pure neutron star—toward larger values of $r$. As $\chi_0$ increases, $\rho_T(r=0)$ decreases and eventually becomes negative for sufficiently large $\chi_0$. By contrast, the cases $\lambda = 1591.55$ and $\lambda = 9.4 \times 10^4$ (middle and right panels) exhibit a qualitatively different behavior. In these regimes, the central value of $\rho_T$ never becomes negative; moreover, for sufficiently large $\chi_0$, $\rho_T(r=0)$ exceeds the central density of the corresponding pure neutron star.  

\begin{figure}[H]
	\begin{centering}
		\includegraphics[scale=.28]{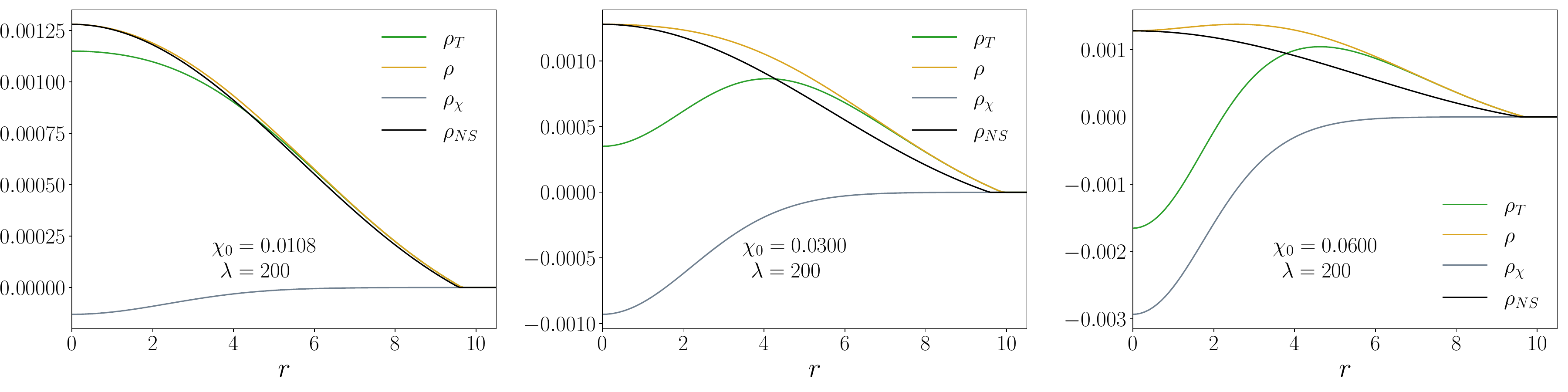}
        \includegraphics[scale=.28]{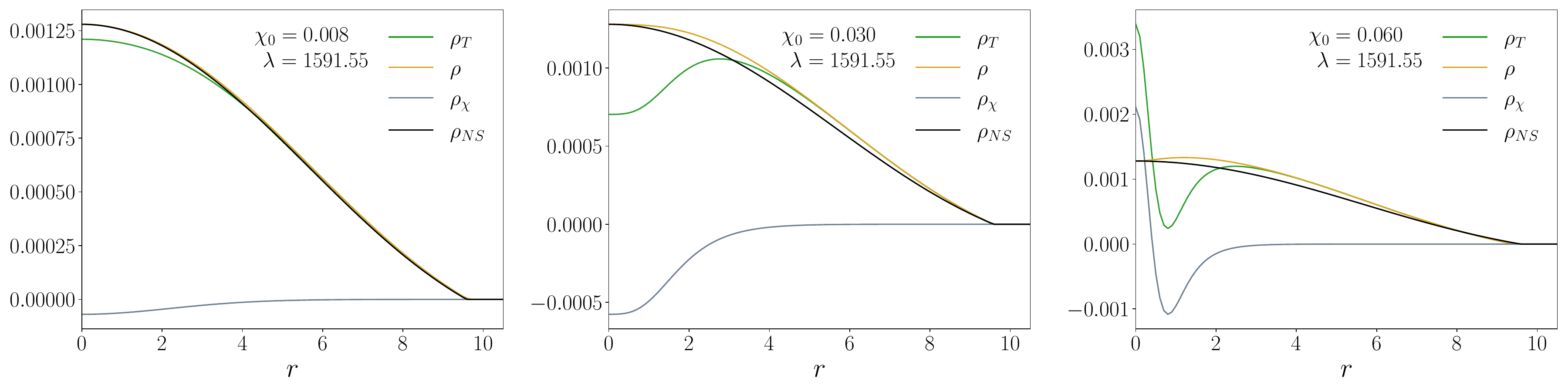}
        \includegraphics[scale=.28]{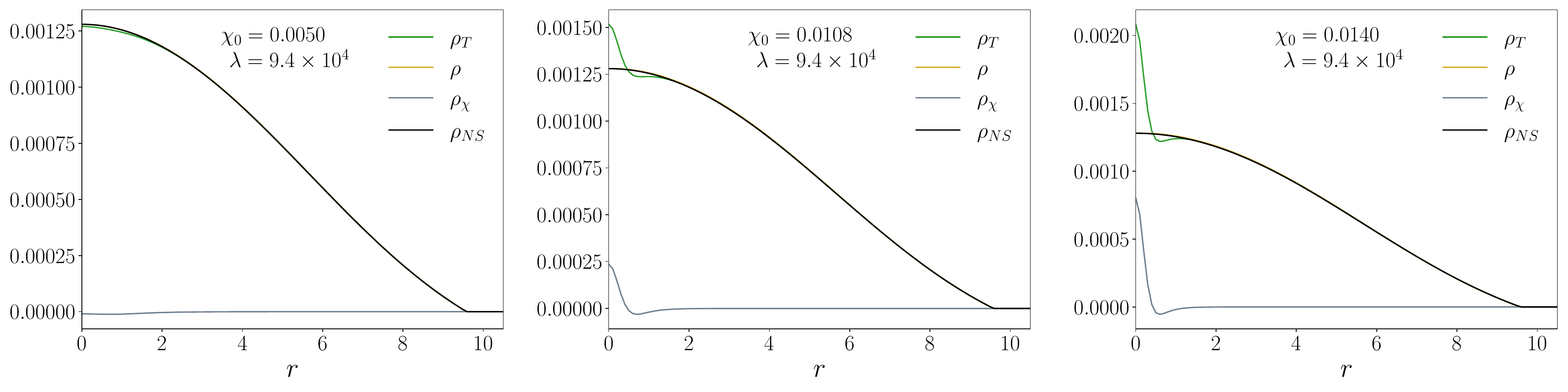}
        \par
        
	\end{centering}
	\caption{Densities profiles for the total density, $\rho_T(r)$, the fermionic density, $\rho(r)$, and phantom density, $\rho_{\chi}(r)$ for configurations with central fermionic density $\rho_0=1.28\times 10^{-3}$, self interaction parameter $\lambda=200$ and three values of the central phantom density: $\chi_0= 0.01080$, left, $\chi_0=0.03$, middle, and $\chi_0=0.06$ right. $\lambda=1591.55$ with $\chi_0 = 0.008, 0.03, 0.06$ and $\lambda=9.4\times10^{4}$, $\chi_0=0.00108,0.01080,0.014$} 
\label{fig:e-NS-L200}
\end{figure}

In Fig.~\eqref{fig:e-NS-L200}, we present the profiles of densities of the ghost scalar field, the fluid, and the total density for three values of the self-interaction constant considered, for the fiducial value of the central density of the neutron star of $\rho_0=0.00128$. 

\begin{figure}[H]
	\begin{centering}
		\includegraphics[scale=.25]{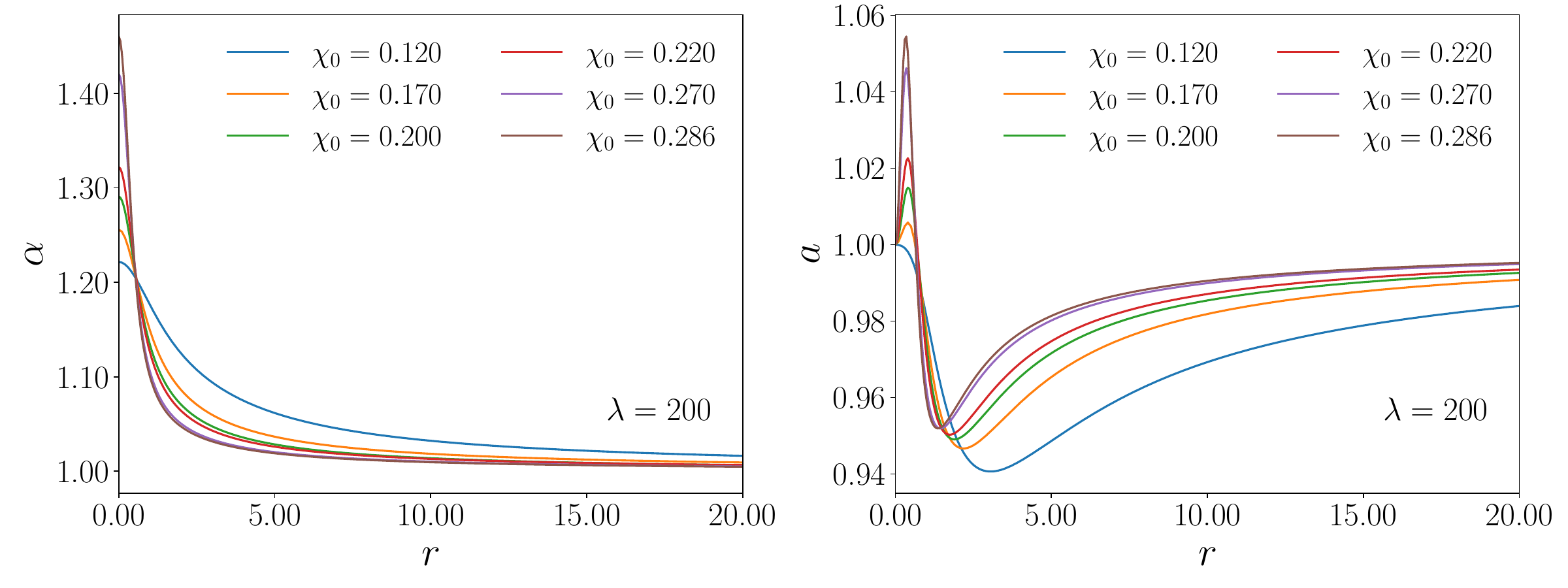}
        \includegraphics[scale=.25]{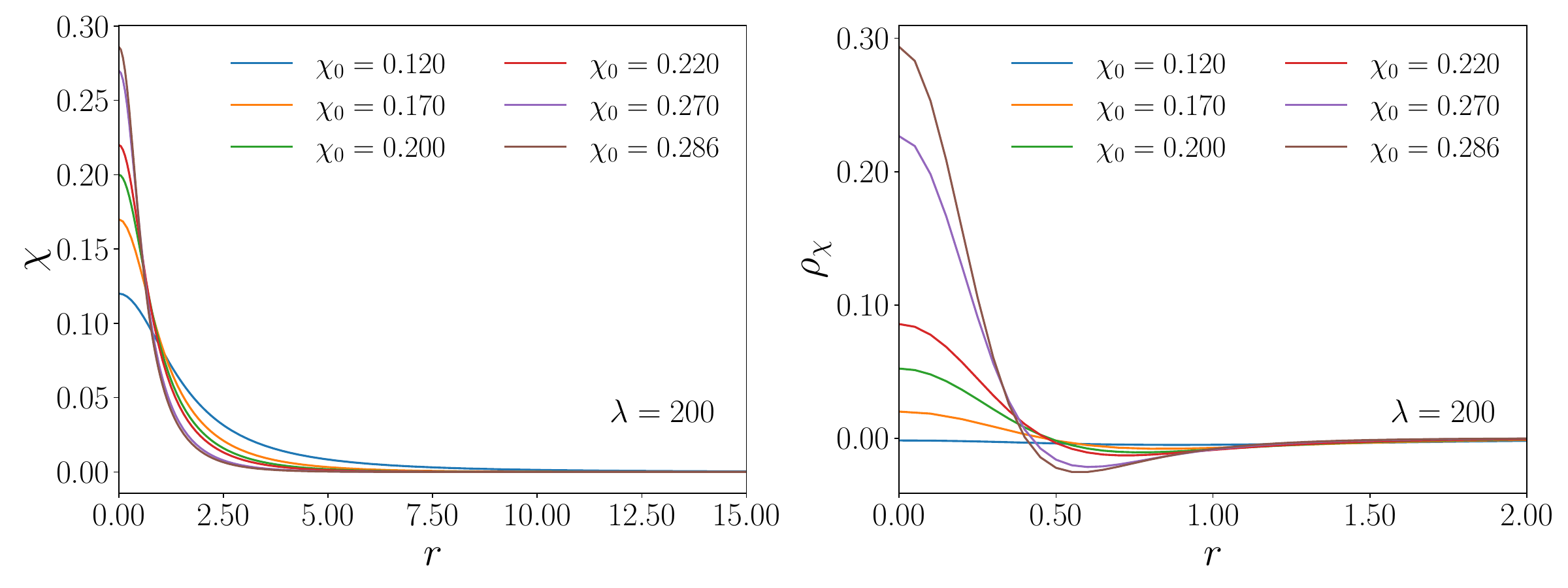}
        \par 
	\end{centering}
	\caption{We present the profiles of the phantom scalar field. The maximum central scalar field $\chi_0$ value corresponds to the real case (frequency equal to 0), whereas the minimum value corresponds to a frequency approaching 1.} 
\label{fig:Phantom}
\end{figure}

Finally, in Fig.~\eqref{fig:Phantom} we present the case of a phantom boson star without being in the gravitational potential of the neutron star: a free phantom (boson) star. We do so for several values of the magnitude of the phantom field at the center, $\chi_0$ for the case of the self interaction parameter , $\lambda=200$. The metric coefficients present a behavior opposite to the one of the usual matter. Notice that the phantom density is positive in the central region and then has the outer region with negative density, approaching zero, from below, as the radius increases.

\section{Basic equations for the evolution}

For the numerical evolutions we adopt a spherically symmetric line element written in isotropic coordinates,
\begin{equation}\label{isotropic_metric}
ds^2 = -\alpha(\hat{r})^2 dt^2 + \psi(\hat{r})^4 \tilde\gamma_{ij} (dx^{i} + \beta^{i}dt)(dx^{j} + \beta^{j}dt),
\end{equation}
where $\alpha$ is the lapse and $\beta^{i}$ the shift vector. The conformal spatial three-metric is given by
\begin{eqnarray}
\tilde\gamma_{ij} dx^i dx^j = a(\hat{r}) d\hat{r}^2 + b(\hat{r}) \hat{r}^2 \left(d\theta^2 + \sin^2\theta d\varphi^2\right) .
\end{eqnarray}
Here, $a(\hat{r})$ and $b(\hat{r})$ denote the isotropic metric functions (not to be confused with other quantities carrying similar notation), $\hat{r}$ is the isotropic radial coordinate used throughout the evolution scheme, and the conformal factor is written as $\psi^4 \equiv e^{4\phi}$. For notational simplicity we replace $\hat{r}\rightarrow r$ in what follows, with the understanding that $r$ always refers to the isotropic radius.

The evolution of the spacetime variables is based on Brown's covariant reformulation~\cite{Brown:2009,Alcubierre:2010is} of the Baumgarte--Shapiro--Shibata--Nakamura (BSSN) system~\cite{nakamura1987general,Shibata95,Baumgarte98}. The evolved fields include the conformal spatial metric $\tilde\gamma_{ij}$, the conformal exponent $\phi$, the trace of the extrinsic curvature $K$, the traceless components $A_a = A^r_{r}$ and $A_b = A^\theta_{\theta} = A^\varphi_{\varphi}$, as well as the radial conformal connection function $\Delta^r$ (see~\cite{Shibata95,Baumgarte98}).

The explicit BSSN evolution equations are not repeated here; they can be found in detail in Ref.~\cite{Montero:2012yr}. We recall that the matter fields enter through projections of the stress--energy tensor $T_{\mu\nu}$: the energy density $\mathcal{E}$, the momentum density $j_i$, and the spatial stress tensor $S_{ij}$, which are defined as
\begin{align}
\mathcal{E} &= n^{\mu} n^{\nu} T_{\mu\nu}, \\
j_i &= -\gamma_i{}^{\mu} n^{\nu} T_{\mu\nu}, \\
S_{ij} &= \gamma_i{}^{\mu} \gamma_j{}^{\nu} T_{\mu\nu}.
\end{align}
In our setup these quantities receive contributions from both the fluid and the complex scalar field. Their explicit forms are provided at the end of this section.

For the gauge choice, we use the non-advective $1+\log$ slicing condition for the lapse $\alpha$, together with a modified Gamma-driver prescription for the shift $\beta^r$. Further discussion of the BSSN equations, gauge conditions, and relativistic hydrodynamics may be found in~\cite{Montero:2012yr}.

We introduce the auxiliary fields
\begin{eqnarray}
\Pi &=& \frac{1}{\alpha}\left(\partial_t - \beta^r \partial_r\right)\chi, \label{Pi_scalar} \\
\Psi &=& \partial_r \chi, \label{Psi_scalar}
\end{eqnarray}
to reduce the Klein--Gordon equation into a first-order system:
\begin{align}
\partial_t \chi &= \beta^r \partial_r \chi + \alpha \Pi, \\
\partial_t \Pi &= \beta^r \partial_r \Pi + \frac{\alpha}{a e^{4\phi}} \biggl[\partial_r \Psi + \Psi \biggl(\frac{2}{r} - \frac{\partial_r a}{2a} + \frac{\partial_r b}{b} + 2\partial_r \phi \biggr)\biggr] + \frac{\Psi}{a e^{4\phi}} + \alpha K \Pi - \alpha\left(\mu^2 - \lambda \chi \bar{\chi}\right)\chi, \\
\partial_t \Psi &= \beta^r \partial_r \Psi + \Psi, \partial_r \beta^r + \partial_r(\alpha \Pi).
\end{align}

The system is supplemented by the Hamiltonian and momentum constraints,
\begin{eqnarray}
\mathcal{H} &=& R - \left(A_a^2 + 2A_b^2\right) + \frac{2}{3}K^2 - 16\pi \mathcal{E} = 0, \label{Hamiltonian constraint} \\
\mathcal{M}_r &=& \partial_r A_a - \frac{2}{3}\partial_r K + 6A_a \partial_r \chi  + (A_a - A_b)\left(\frac{2}{r} + \frac{\partial_r b}{b}\right) - 8\pi j_r = 0, \label{Momentum constraint}
\end{eqnarray}
where $R$ denotes the Ricci scalar.

The scalar-field (bosonic) contributions to the matter source terms are 
\begin{eqnarray}
\mathcal{E}^{\rm SF} &=&- \frac{1}{2}\left(\bar{\Pi} \Pi + \frac{\bar{\Psi}\Psi}{e^{4\phi} a}\right) - \frac{1}{2}\mu^2 \bar{\chi}\chi + \frac{1}{4}\lambda (\bar{\chi}\chi)^2, \label{rhomat_phi}\\
j_r^{\rm SF} &=& +\frac{1}{2}\left(\bar{\Pi}\Psi + \bar{\Psi}\Pi\right), \label{j_phi}\\
S_a^{\rm SF} &=& -\frac{1}{2}\left(\bar{\Pi}\Pi + \frac{\bar{\Psi}\Psi}{e^{4\phi}a}\right) + \frac{1}{2}\mu^2 \bar{\chi}\chi - \frac{1}{4}\lambda (\bar{\chi}\chi)^2, \\
S_b^{\rm SF} &=& -\frac{1}{2}\left(\bar{\Pi}\Pi - \frac{\bar{\Psi}\Psi}{e^{4\phi}a}\right) + \frac{1}{2}\mu^2 \bar{\chi}\chi - \frac{1}{4}\lambda (\bar{\chi}\chi)^2,
\end{eqnarray}
where $S_a = {S^r}_r$ and $S_b = {S^\theta}_\theta = {S^\varphi}_\varphi$. The fluid (fermionic) contributions are
\begin{align}
\mathcal{E}^{\rm fluid} &= \left[\rho(1+\epsilon) + P\right] W^2 - P, \\
j_r^{\rm fluid} &= e^{4\phi} a \left[\rho(1+\epsilon) + P\right] W^2 v^r, \\
S_a^{\rm fluid} &= e^{4\phi} a \left[\rho(1+\epsilon) + P\right] W^2 v^r + P, \\
S_b^{\rm fluid} &= P,
\end{align}
with $W = \alpha u^t$ the Lorentz factor and $v^r$ the radial component of the fluid three-velocity.

%%%%%%%%%%%%%%%%%%%%%%%%%%%%%%%%%%%%%%%%%%%%%%%%%%%%
\section{Numerical framework} 
\label{sec:numerics}
%%%%%%%%%%%%%%%%%%%%%%%%%%%%%%%%%%%%%%%%%%%%%%%%%%%%

The Einstein–Klein–Gordon–Euler system is evolved using a numerical-relativity framework first introduced by~\cite{Montero:2012yr} and later extended to incorporate the dynamics of fundamental bosonic fields~\cite{Sanchis-Gual:2014ewa,Sanchis-Gual:2015sxa,Sanchis-Gual:2015lje,Escorihuela-Tomas:2017uac,Sanchis-Gual:2017bhw,DiGiovanni:2018bvo,DiGiovanni:2020frc}.

To integrate the evolution equations in time, we adopt the Partially Implicit Runge–Kutta (PIRK) method developed in~\cite{Cordero-Carrion:2012qac,cordero2014partially}.  This scheme handles numerical instabilities generated by the $1/r$ terms. Whereas equilibrium solutions are constructed in Schwarzschild coordinates on a uniform linear radial grid, the dynamical simulations employ isotropic coordinates together with a logarithmic radial discretization. Specifically, the computational domain is covered by an isotropic grid consisting of two regions: an inner segment following a geometric progression and an outer region using a hyperbolic-cosine distribution. This setup enables the outer boundary to be placed at a sufficiently large radius, minimizing spurious reflections. Additional details on the grid structure are given in~\cite{Sanchis-Gual:2015sxa}.

Our simulations use a minimum radial resolution of $\Delta r = 0.05$, with the inner boundary located at $r_{\rm min} = \Delta r/2$ and the outer boundary at $r_{\rm max} = 6000$. The timestep is fixed as $\Delta t = 0.3\Delta r$, which ensures stable long-term evolutions. To suppress high-frequency numerical artifacts, 4th-order Kreiss–Oliger dissipation is applied, and advection terms (e.g., $\beta^r \partial_r f$) are discretized using an upwind scheme. Radiative (Sommerfeld) boundary conditions are imposed at the outer edge of the domain.

\section{Evolution} \label{sec:evolution}

As discussed in section \ref{Sec:Initial_c}, neutron stars can shelter a wide range of nuclei of ghost matter, constituting a small fraction of their density, without provoking important changes in their configuration. Even though comparatively small, it is remarkable that ghost matter, characterized by its dispersive nature, can be confined within an effective finite volume due to the gravitational well of neutron stars.

\begin{figure}[H]
	\begin{centering}
  \includegraphics[scale=.42]{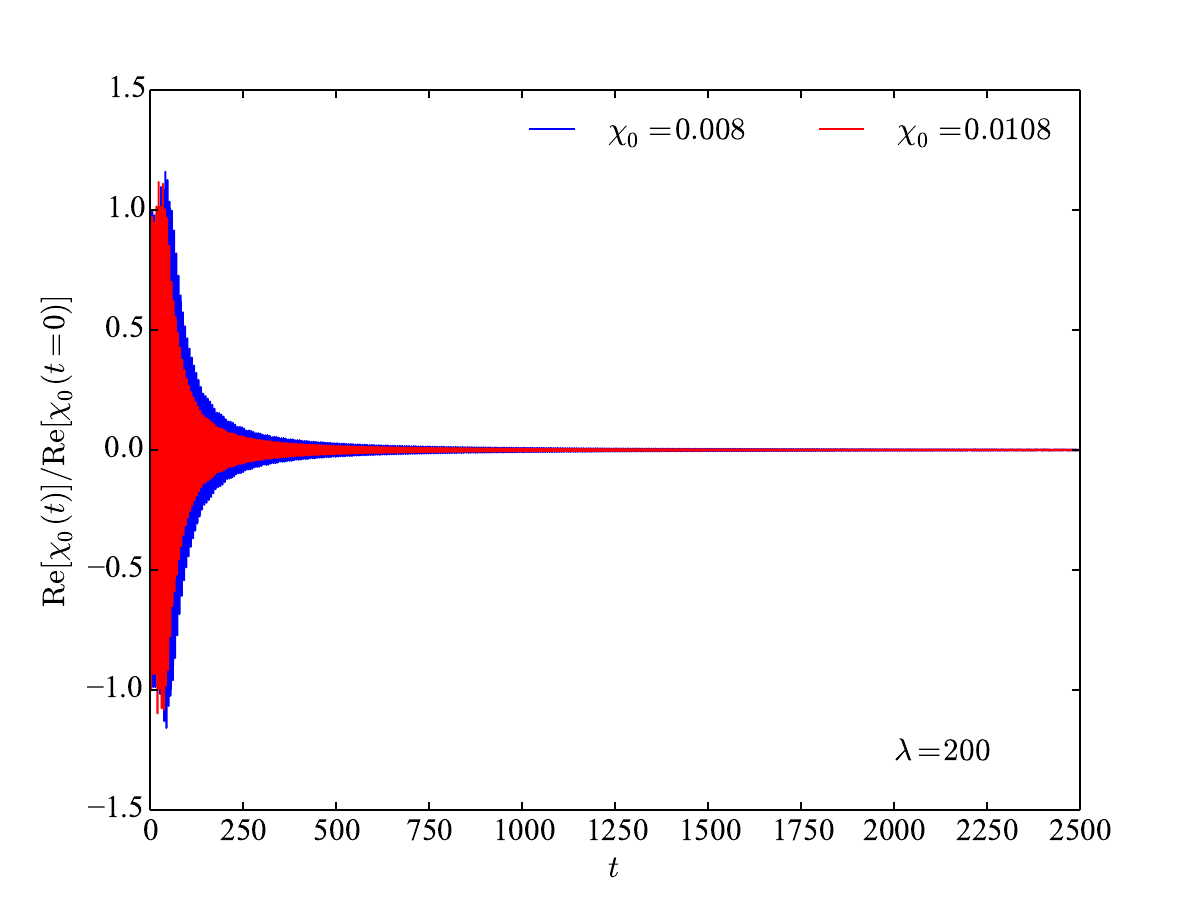}  \includegraphics[scale=.42]{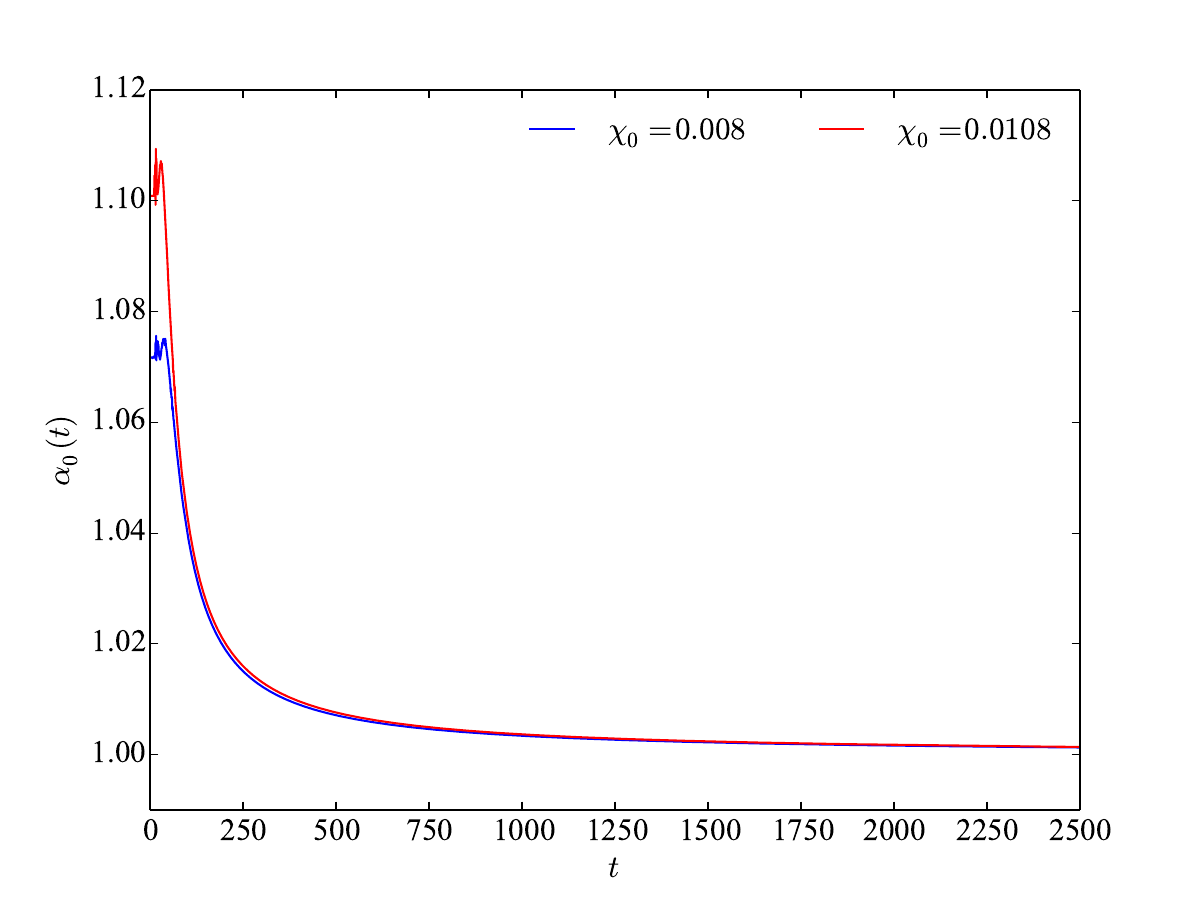}
		\par
	\end{centering}
	\caption{(Left panel) Time evolution of the real part of the scalar field extracted at $r_{\rm obs}=5$ for different values of the initial $\chi_0=\lbrace0.008,0.0108\rbrace$ for the ghost boson star case. (Right panel) Time evolution of the central value of the lapse $\alpha$.}
\label{fig:ghost-r}
\end{figure}

The question that arises is if these models are dynamically viable. We proceed to study some illustrative initial configurations solving the fully non-linear Einstein-Euler-(ghost) Klein-Gordon system. First, we start by considering ghost scalar field configurations without a confining neutron star: ghost boson stars. In Fig.~\ref{fig:ghost-r}, we plot the evolution of the amplitude of the real part of the scalar field at a given radius $r_{\rm obs}=5$ and the central value of the lapse for the cases with $\chi_0=\lbrace0.008,0.0108\rbrace$. As expected, their fate is clear: a single ghost boson star cannot form bound states and quickly becomes unstable, dispersing away. The central lapse returns to 1 and the scalar field amplitude in the inner region goes to zero.

We now focus on mixed configurations containing both fermionic and bosonic components. We evolve mixed stars with an initial central fermion rest-mass density $\rho_0=1.28\times10^{-3}$ and different scalar-field configurations characterized by the central amplitude $\chi_0$, considering three values of the self-interaction parameter $\lambda=\lbrace200,1591.55,90000\rbrace$. All simulations are evolved up to a final time $t_{\rm final}=6000$. Our numerical results show that several of these mixed stars remain stable and, moreover, that a remarkable synchronization phenomenon emerges between the oscillations of the neutron star and those of the scalar field~\cite{lazarte2025gravitational}.

\begin{figure}[H]
	\begin{centering}
		\includegraphics[scale=0.36]{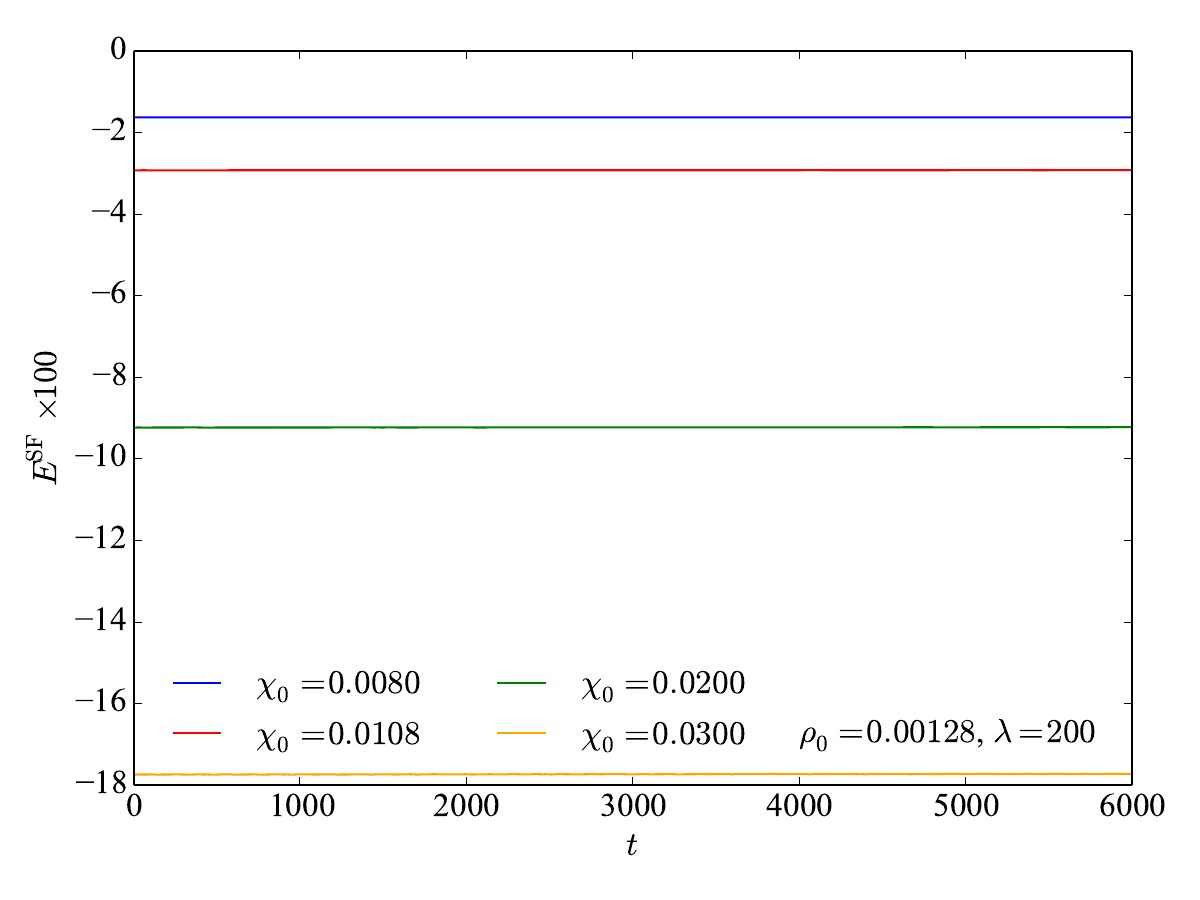}\includegraphics[scale=0.36]{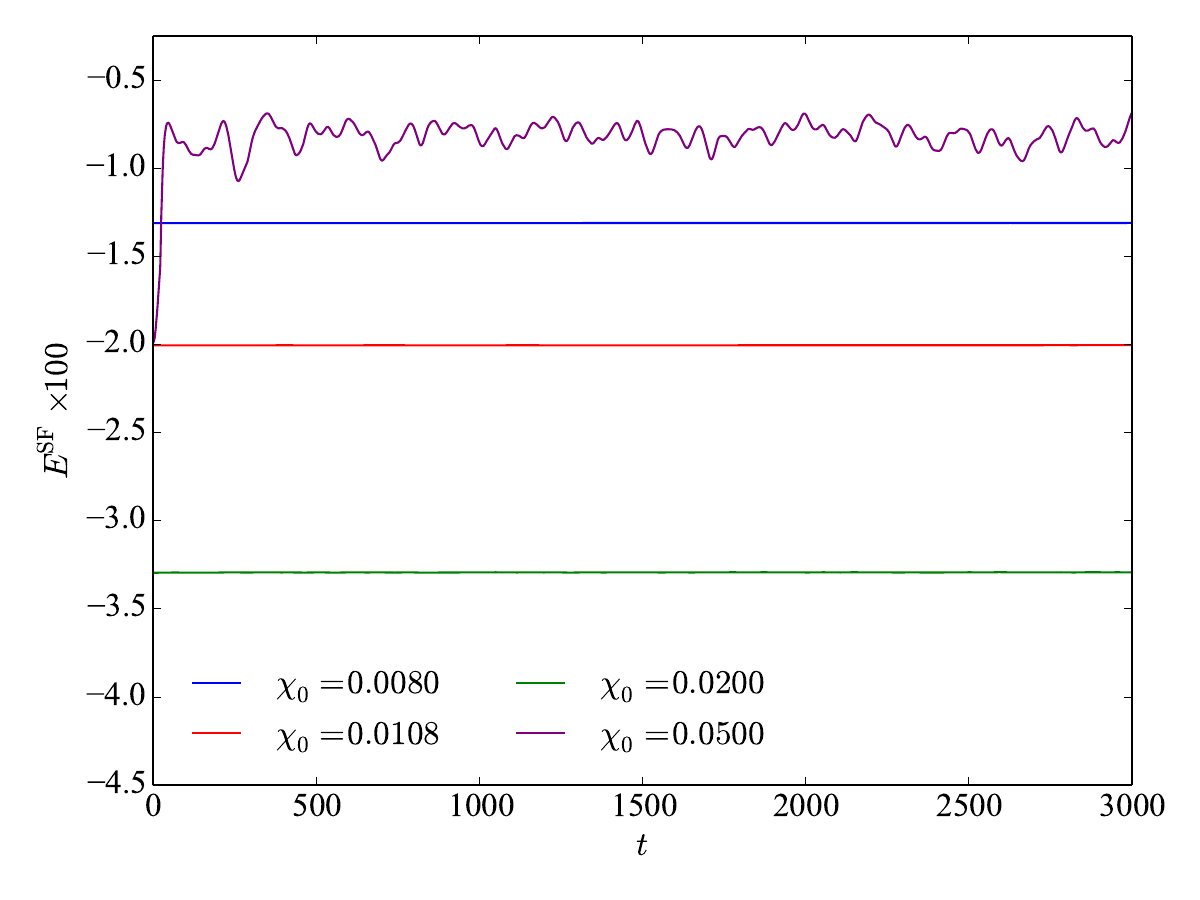}
        \includegraphics[scale=0.36]{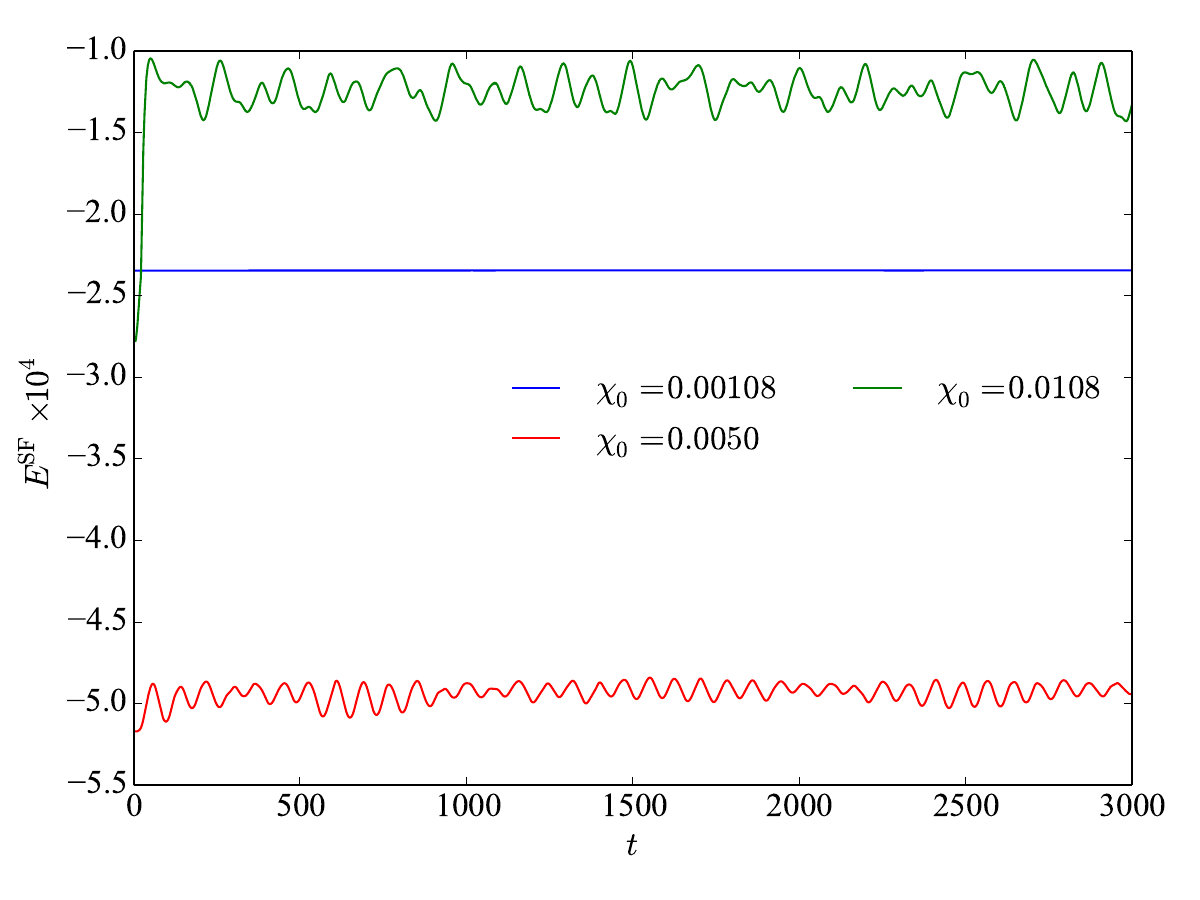}
		\par
	\end{centering}
	\caption{Time evolution of the energy of the scalar field computed up to $r_{*}=30$ for different initial values of $\chi_0$ and $\lambda=200$ (top left panel), $\lambda=1591.55$ (top right panel), and $\lambda=90000$ (bottom panel).}
\label{fig:energy_scalar}
\end{figure}

In Fig.~\ref{fig:energy_scalar} we show the volume integral of the scalar-field energy density, $\mathcal{E}^{\rm SF}$, given by Eq.~\eqref{rhomat_phi}, up to a radius $r_{}$, defined as
\begin{equation}
E^{\rm SF}=\int_0^{r_{}} dV\,\mathcal{E}^{\rm SF}.
\end{equation}

Starting with the top-left panel, corresponding to $\lambda=200$ and $\chi_0=\lbrace0.0080,0.0108,0.0200,0.0300\rbrace$, we observe that the scalar-field energy remains approximately constant throughout the evolution, with variations below $1\%$, indicating stable configurations. Similar behavior is found for the other two values of $\lambda$ (top-right and bottom panels). For $\lambda=1591.55$, the configurations remain stable up to $\chi_0\sim0.030$, while for $\lambda=90000$ stability is achieved only for smaller amplitudes, down to $\chi_0=0.00108$. For larger values of $\chi_0$, the system becomes unstable and sheds part of the scalar field. In these cases, the scalar-field energy decreases in absolute value due to partial dispersal, after which the system relaxes toward a new equilibrium characterized by a smaller $|E^{\rm SF}|$.

\begin{figure}[t]
	\begin{centering}
		\includegraphics[scale=.36]{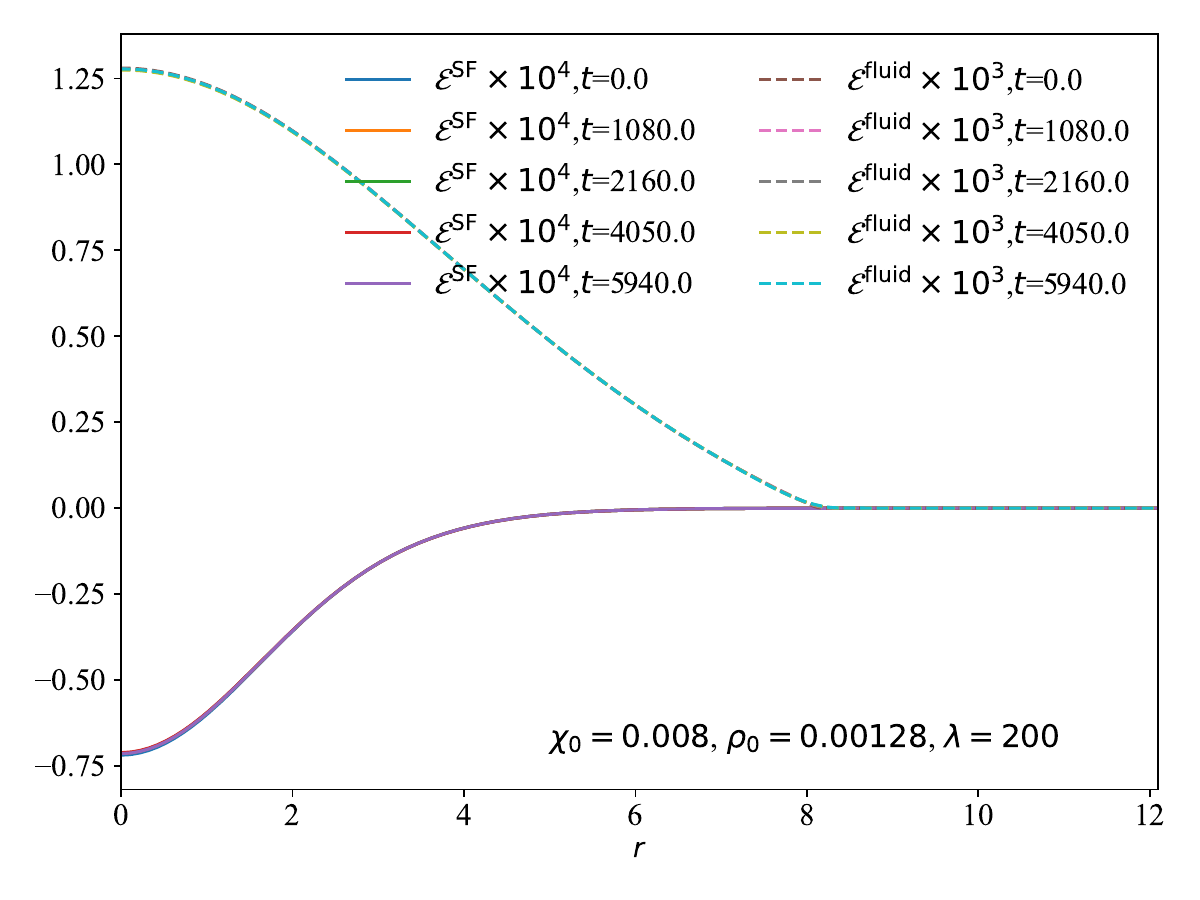} \hspace{1 cm}
  \includegraphics[scale=.36]{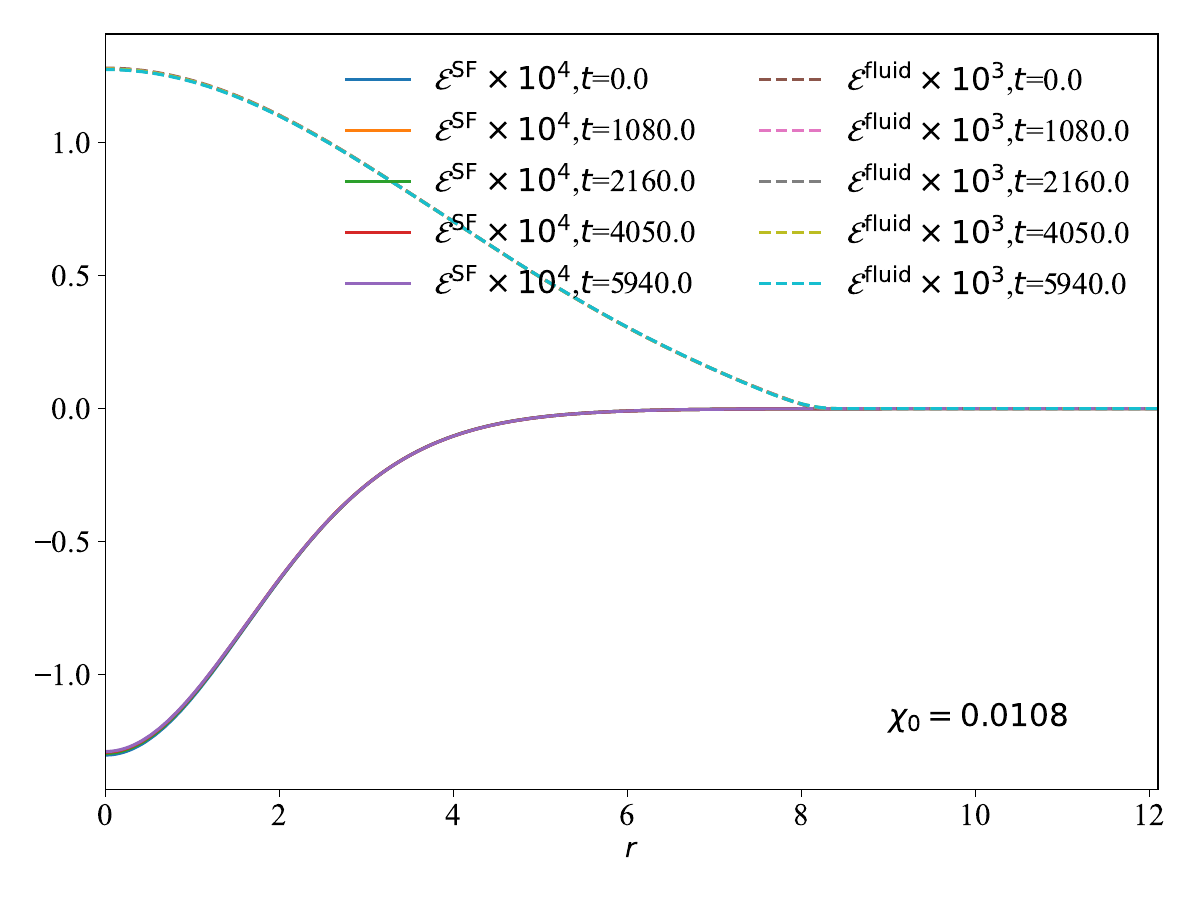} \\
  \includegraphics[scale=.36]{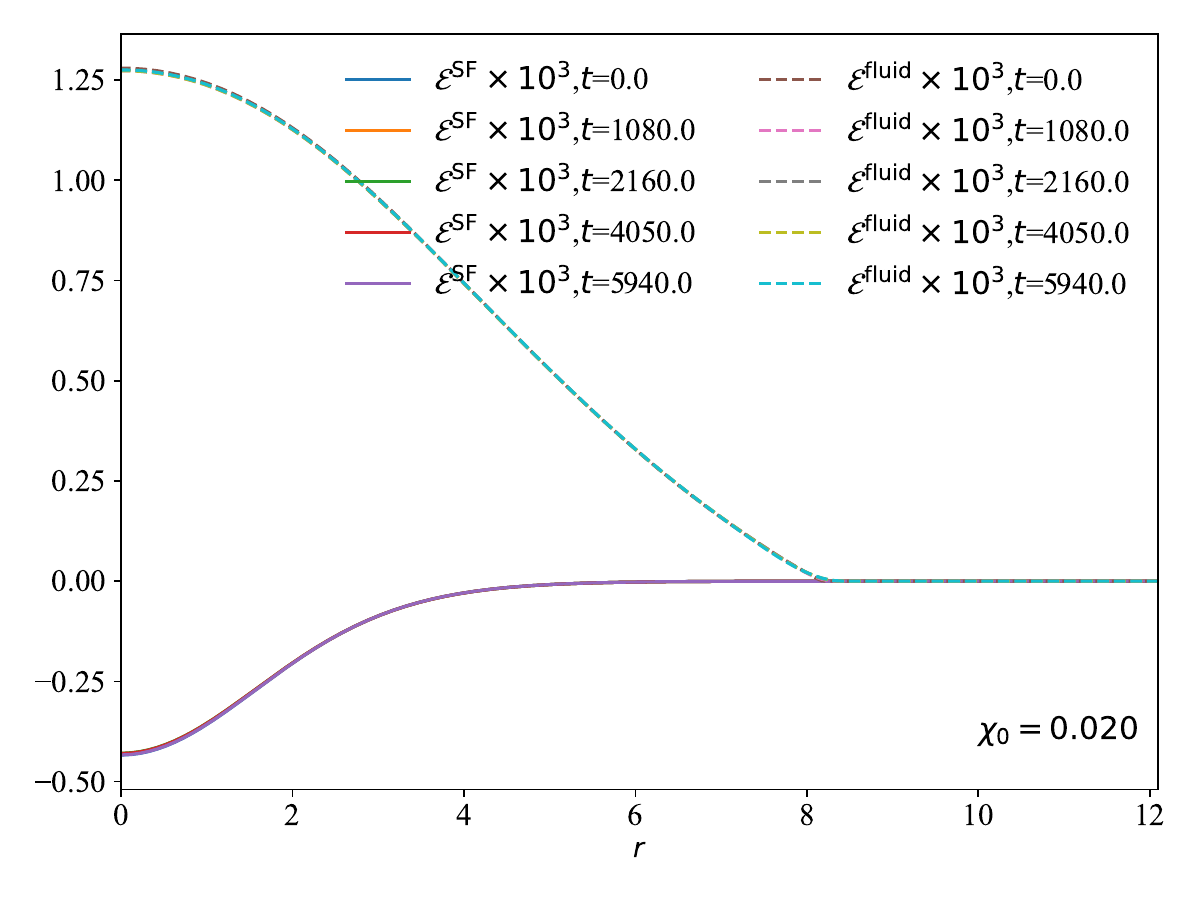} \hspace{1 cm}
  \includegraphics[scale=.36]{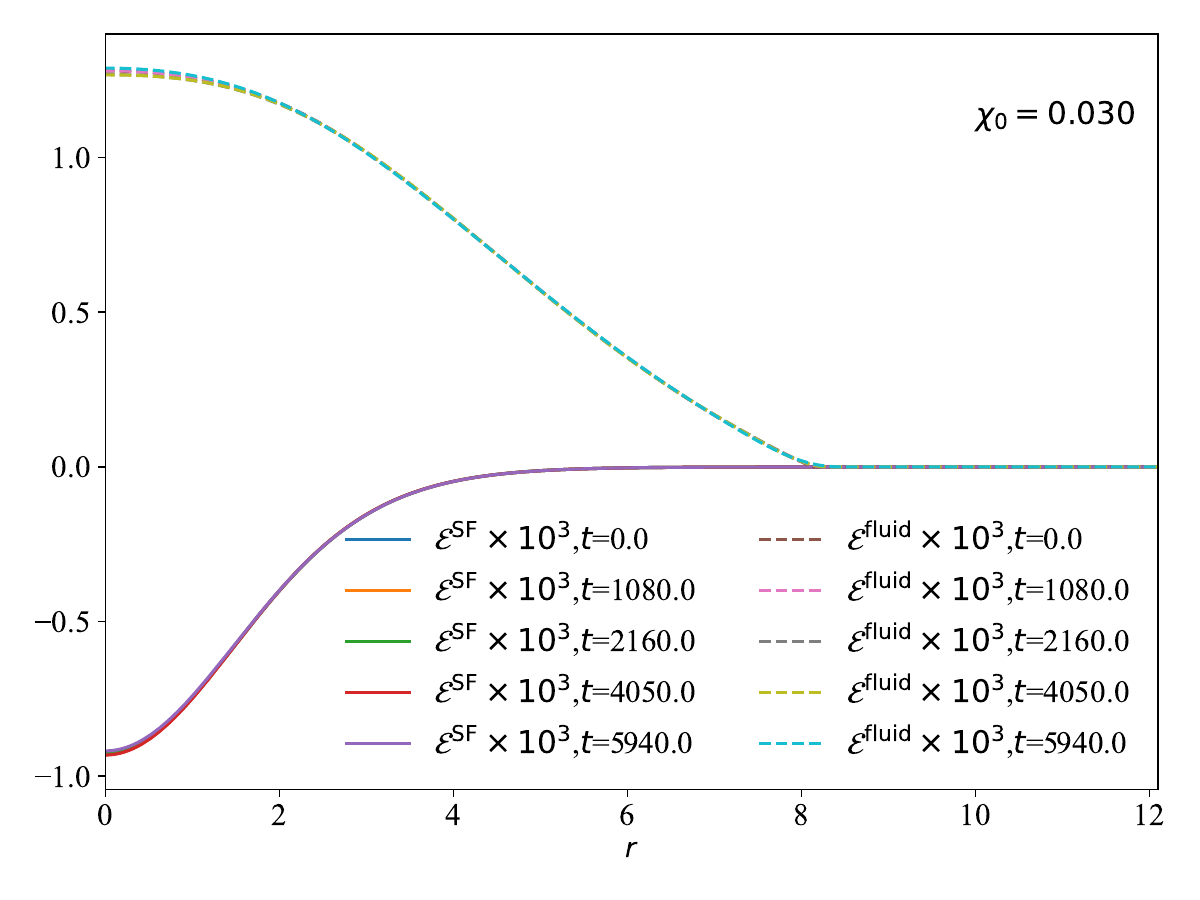}
		\par
	\end{centering}
	\caption{Snapshots of the time evolution of the energy density $\varepsilon$ of scalar component and the fluid rest-mass density for the mixed stars with $\rho_0=1.28\times10^{-3}$ and $\chi_0=\lbrace0.008,0.0108,0.020,0.030\rbrace$ for $\lambda=200$. The configurations exhibit dynamical stability throughout the evolution.}
\label{fig:eps-evol1}
\end{figure}

\begin{figure}[t]
	\begin{centering}
		\includegraphics[scale=.36]{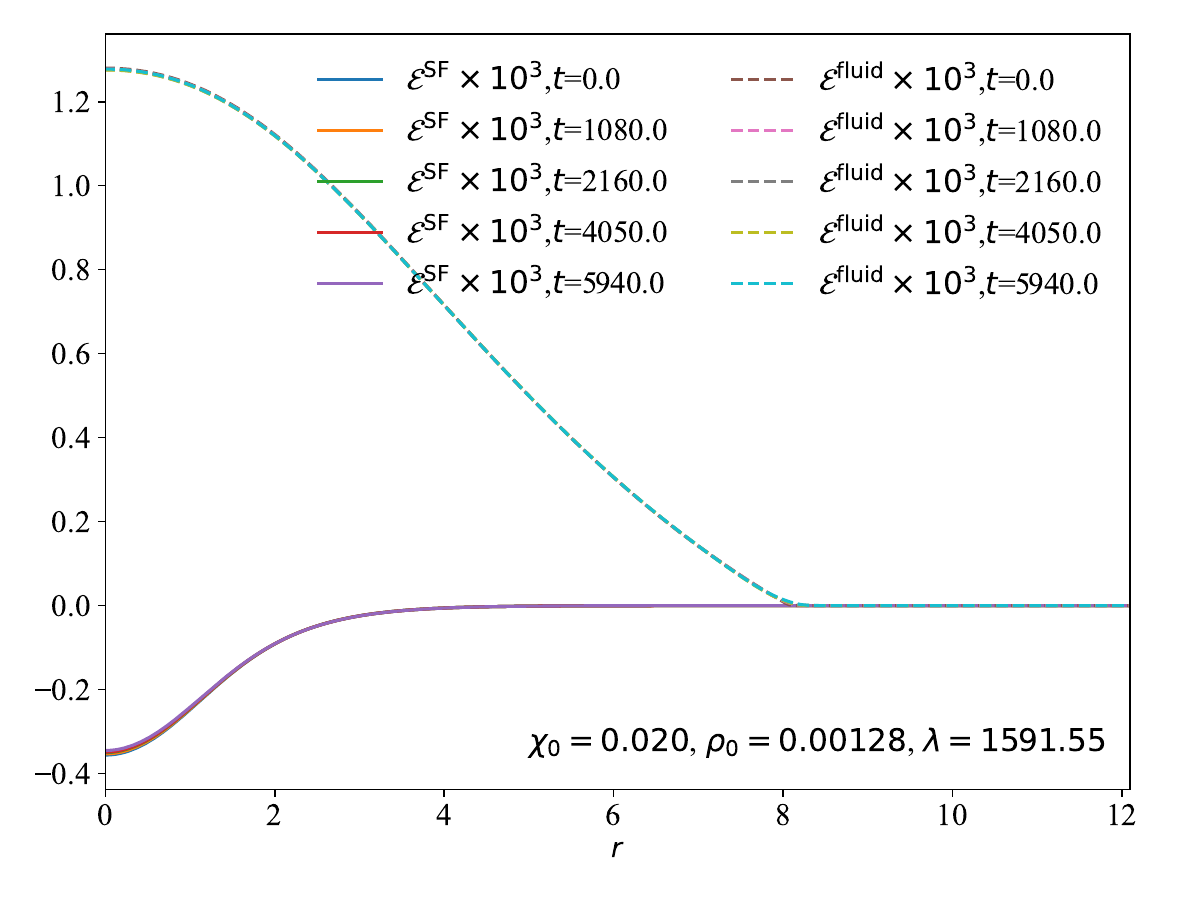} \hspace{1 cm}
  \includegraphics[scale=.36]{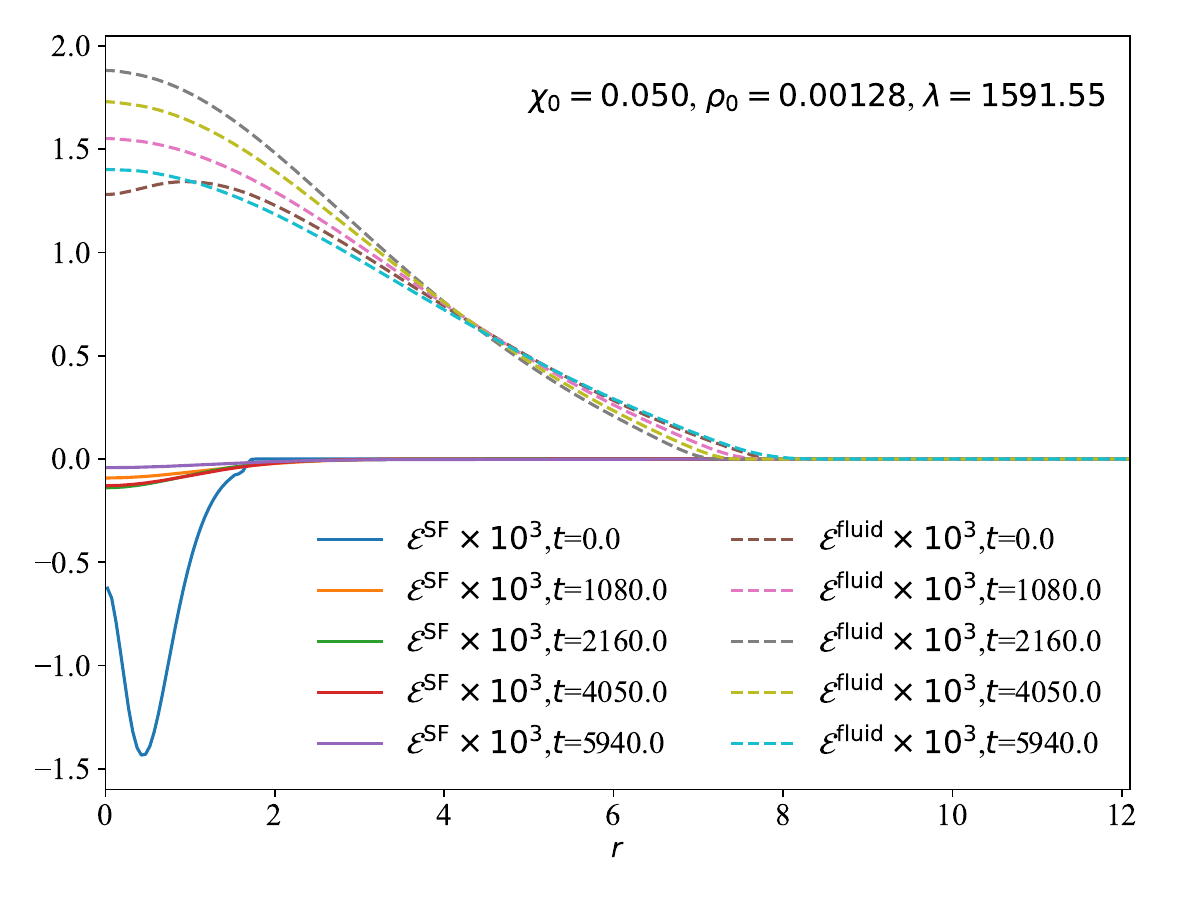}
		\par
	\end{centering}
	\caption{Snapshots of the time evolution of the energy density $\varepsilon$ of scalar component and the fluid rest-mass density  for the mixed stars with $\rho_0=1.28\times10^{-3}$ and $\chi_0=\lbrace0.020,0.050\rbrace$ for $\lambda=1591.55$. In this case, for the larger initial value of the central ghost field, $\chi_0=0.050$, notice how a large fraction of the ghost scalar field disperses and modifies the neutron star density profile.}
\label{fig:eps-evol2}
\end{figure}

In Figs.~\eqref{fig:eps-evol1} and~\eqref{fig:eps-evol2} we show the radial profiles of the scalar-field energy density and the fermion rest-mass density at different times for models with $\lambda=200$ and $\lambda=1591.55$. For $\lambda=200$, we observe that the bosonic component remains confined within the neutron star for all considered values of the initial scalar-field amplitude $\chi_0$ up to $t_{\rm final}=6000$, confirming the stability of these configurations.

For the larger self-interaction parameter $\lambda=1591.55$, we instead confirm the onset of instabilities for configurations with sufficiently large $\chi_0$. In these cases, part of the scalar field disperses, as illustrated in the right panel of Fig.~\ref{fig:eps-evol2}, corresponding to $\chi_0=0.050$. During this process, both the scalar energy and the fermion rest-mass density profiles evolve dynamically due to the loss of a fraction of the bosonic component. However, a remnant scalar field remains gravitationally bound within the neutron star throughout the simulation. The system relaxes toward a new stable mixed configuration characterized by a smaller scalar-field amplitude.

\begin{figure}[t]
	\begin{centering}
		\includegraphics[scale=0.36]{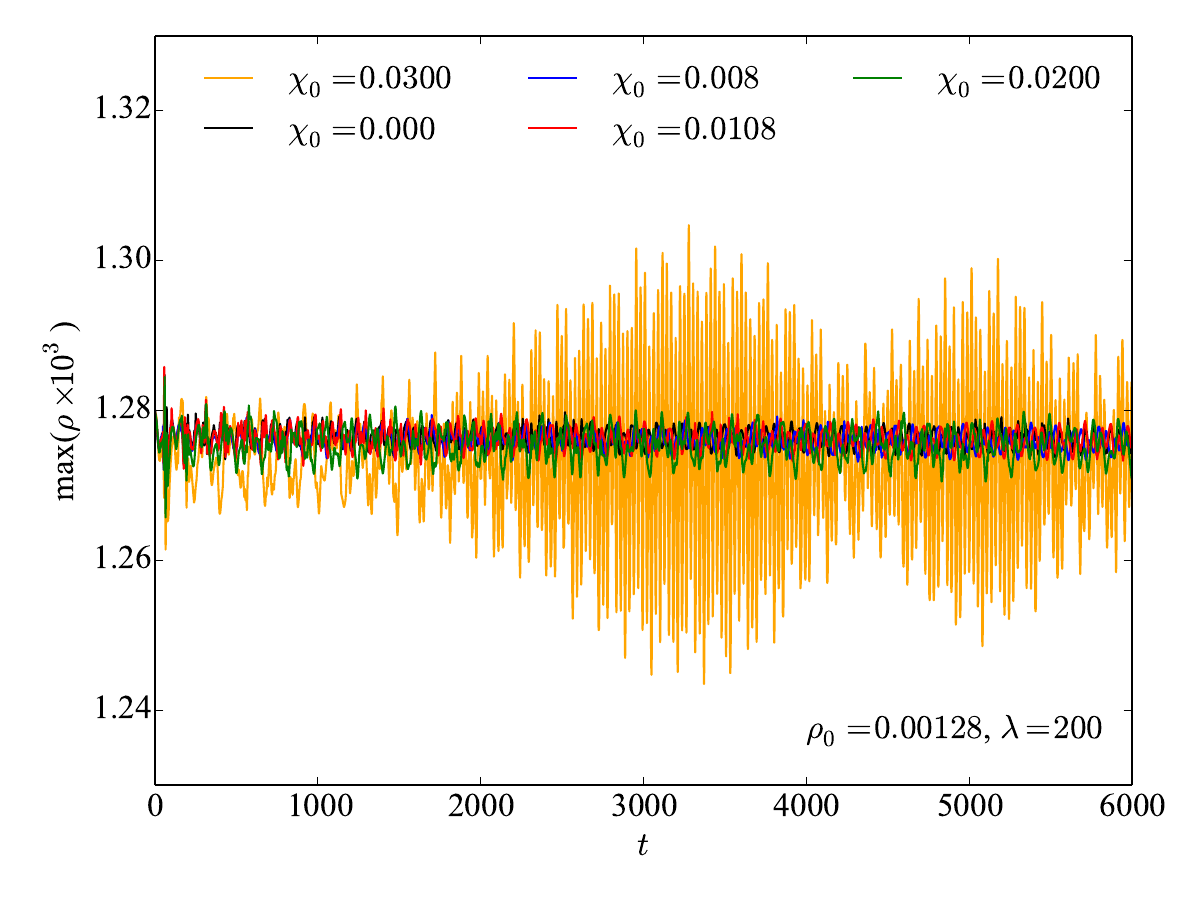}\includegraphics[scale=0.36]{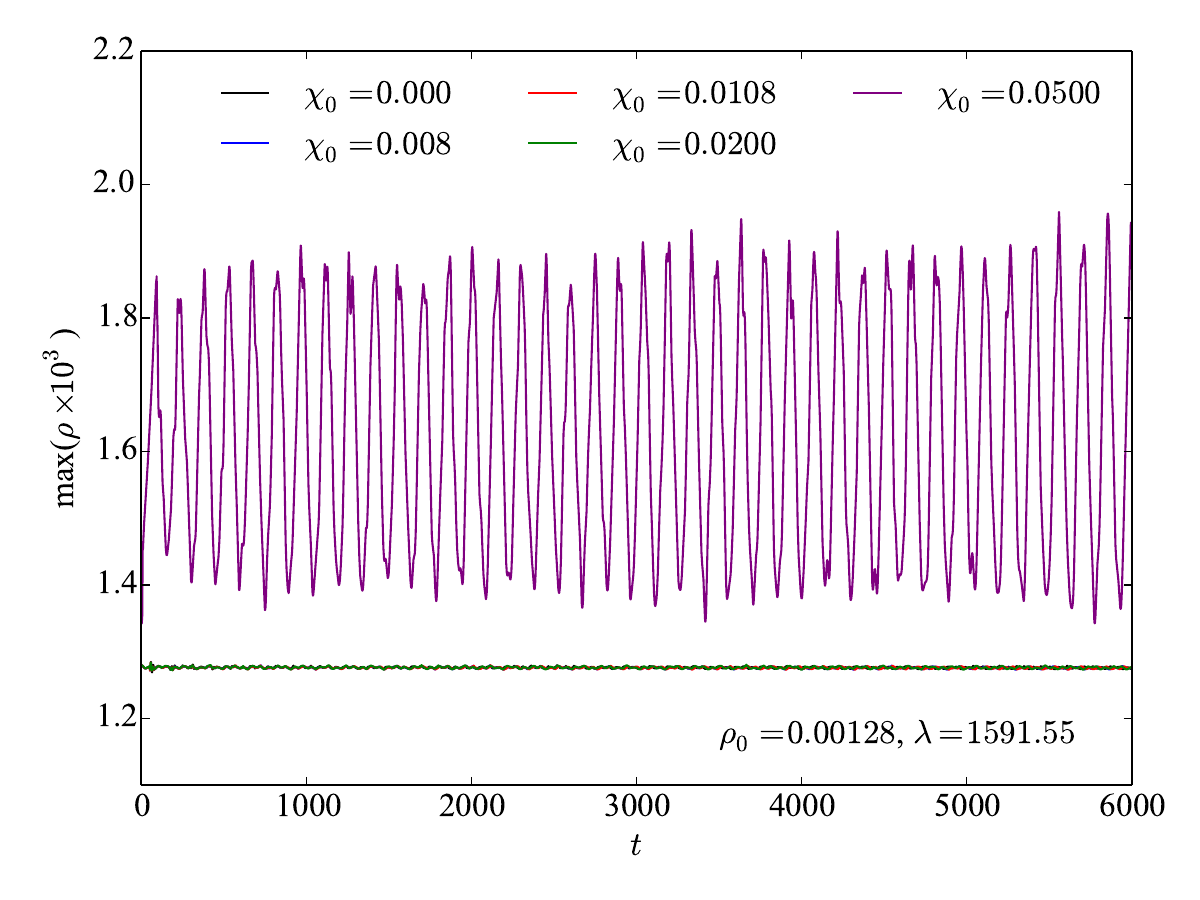}
        \includegraphics[scale=0.36]{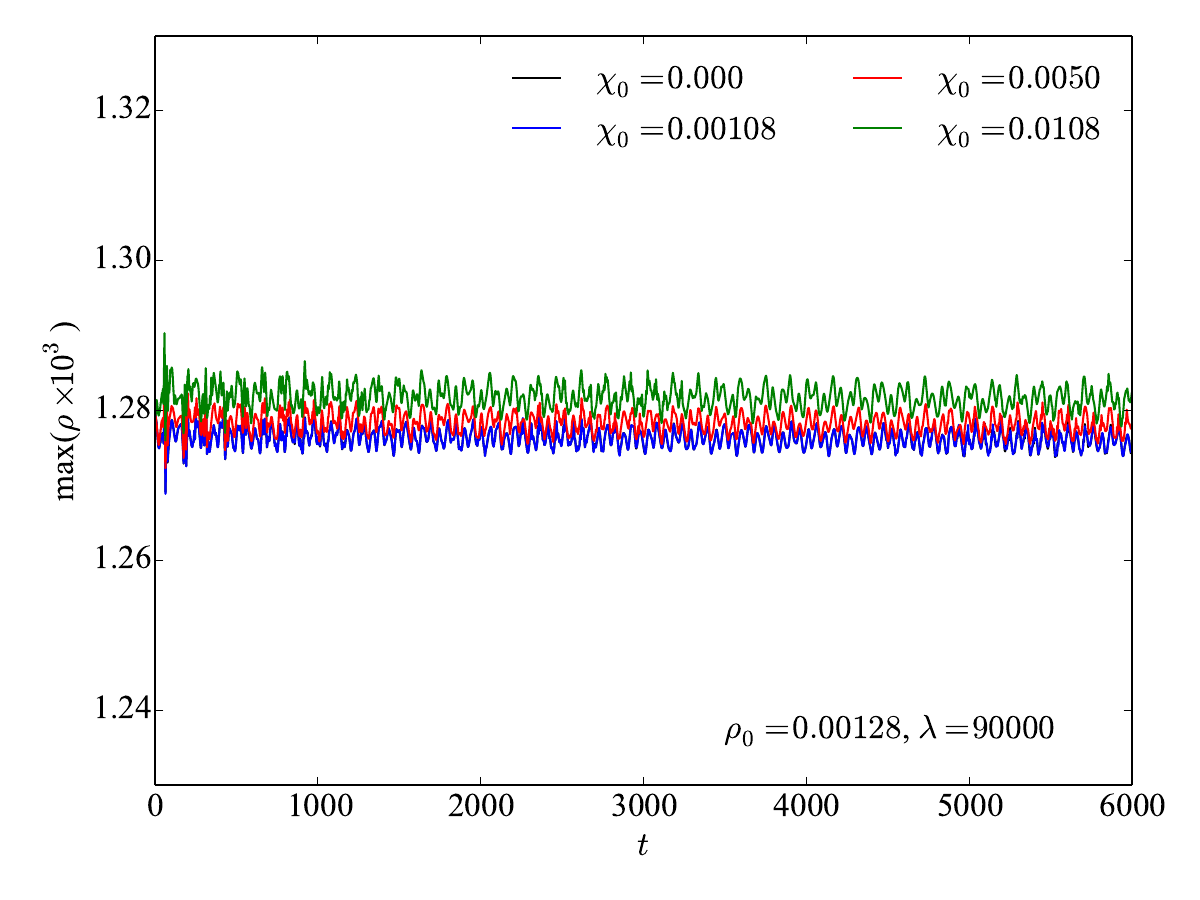}
		\par
	\end{centering}
	\caption{Time evolution of the maximum value of the rest-mass density of the fluid for different initial values of $\chi_0$ and $\lambda=200$ (top left panel), $\lambda=1591.55$ (top right panel), and $\lambda=90000$ (bottom panel).}
\label{fig:OliM_1}
\end{figure}

In contrast, for the stable configurations, small perturbations arising from numerical truncation errors induce only mild oscillations in both the scalar field and the neutron star, while the overall structure of the system remains essentially unchanged.
Fig.~\eqref{fig:OliM_1} displays the time evolution of the central rest-mass density of the neutron star for several mixed configurations with $\lambda=\lbrace200,1591.55,9000\rbrace$. The oscillations are driven by both numerical truncation errors and the presence of the scalar field. As the scalar-field content increases, the amplitude of the induced radial pulsations grows accordingly. However, for the stable configurations, these oscillations remain very small (within $\sim1$--$2\%$ of the initial value), indicating that the fermionic component remains dynamically stable.

The top-right and bottom panels of Fig.~\eqref{fig:OliM_1} correspond to unstable configurations. In these cases, part of the scalar field is expelled from the system, causing the neutron star to evolve toward a more compact configuration (see the right panel of Fig.~\eqref{fig:eps-evol2}) and leading to a noticeable increase in the central rest-mass density.

\begin{figure}[H]
	\begin{centering}
		\includegraphics[scale=.42]{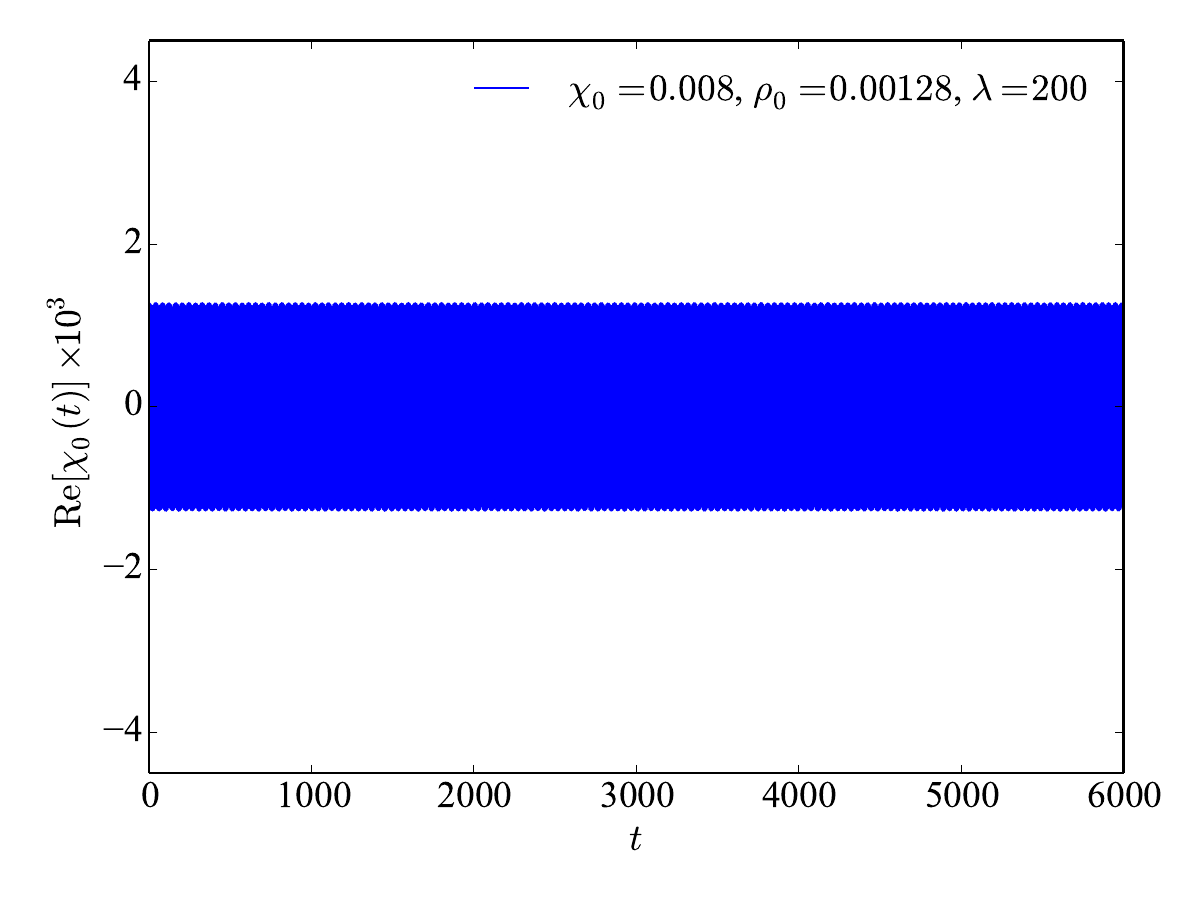}
  \includegraphics[scale=.42]{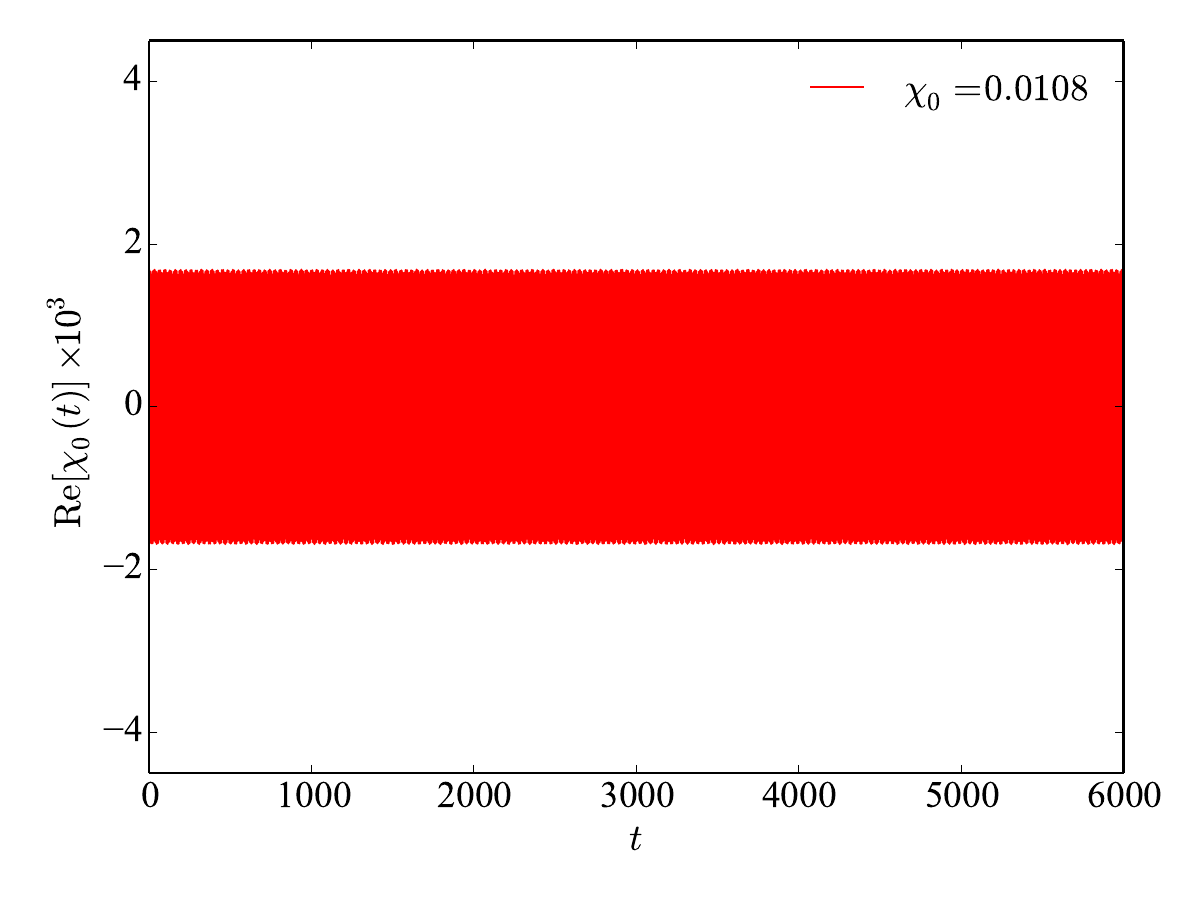}
  \includegraphics[scale=.42]{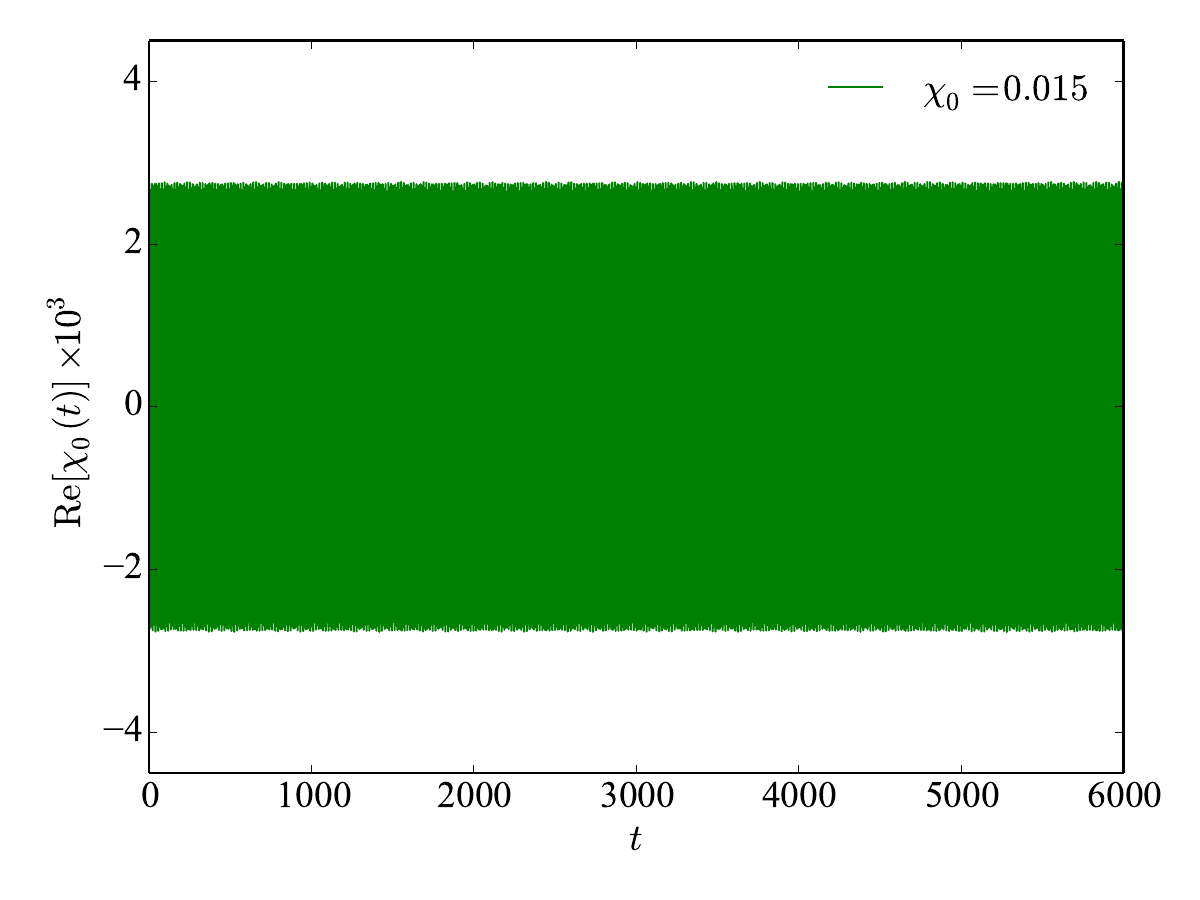}
  \includegraphics[scale=.42]{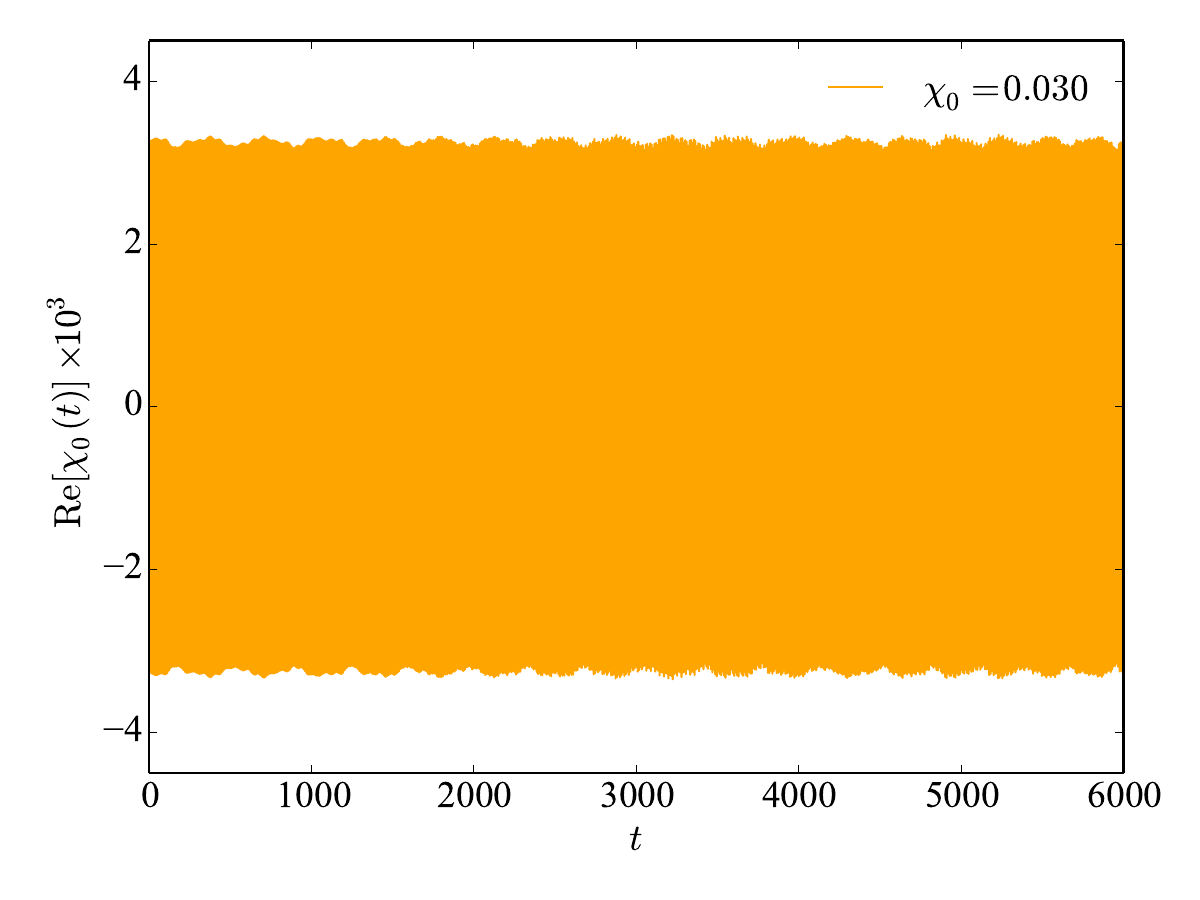}
		\par
	\end{centering}
	\caption{Time evolution of the real part of the scalar field extracted at $r_{\rm obs}=5$ for different values of the initial $\chi_0$ and $\lambda=200$. }
\label{fig:OliM_2}
\end{figure}

As expected of a complex field, the real and imaginary parts of the scalar field oscillate in time with a characteristic frequency $\omega$, inherited from the harmonic time dependence of the field, even though the energy density is initially time-independent. For a stable configuration, the scalar-field amplitude should remain constant in time. Figure~\eqref{fig:OliM_2} displays the evolution of the real part of the scalar field with $\lambda=200$ (the imaginary part behaves identically). For all $\chi_0$, we do not observe any noticeable change. Such a change would correspond to the decay into a less massive bosonic component and thus produce a shift in its fundamental frequency~$\omega$.

\begin{figure}[H]
	\begin{centering}
		\includegraphics[scale=0.4]{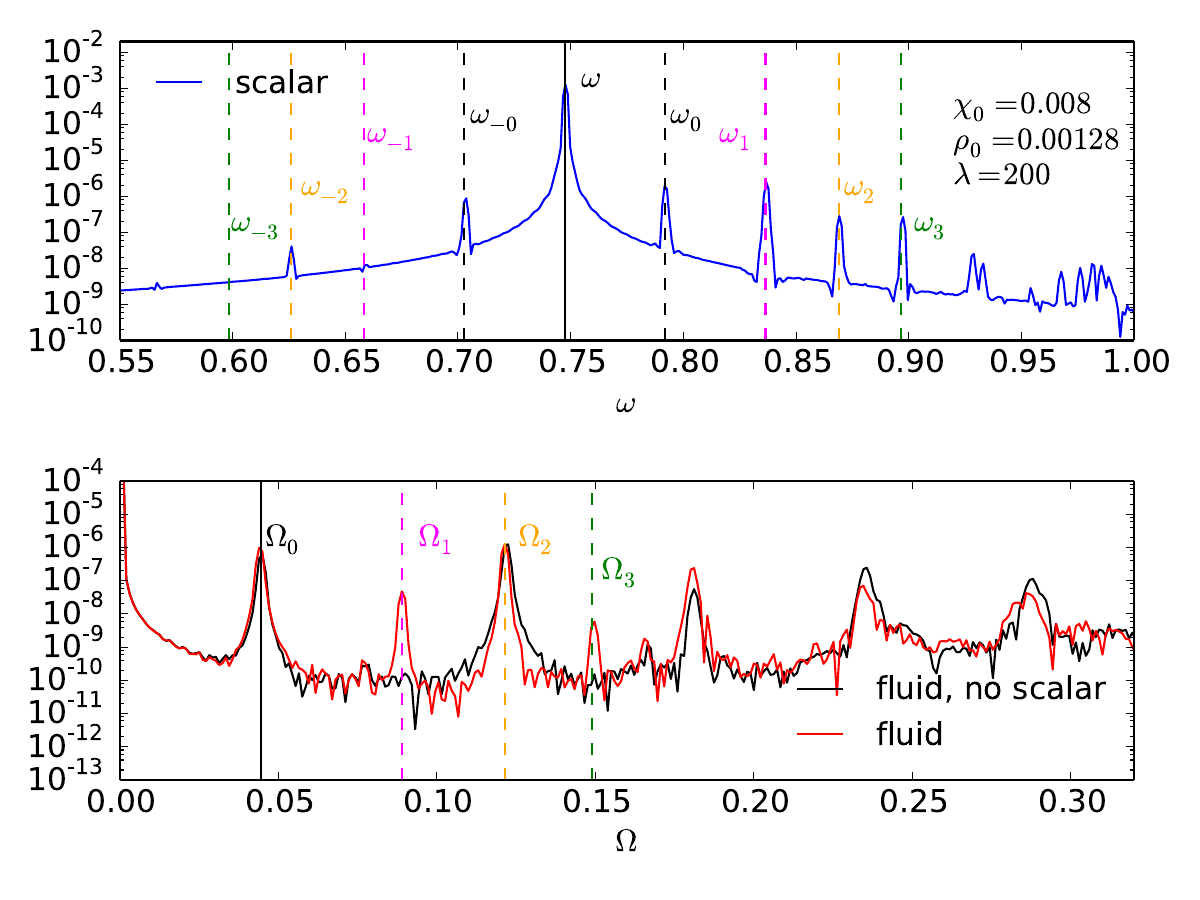}
        \includegraphics[scale=0.4]{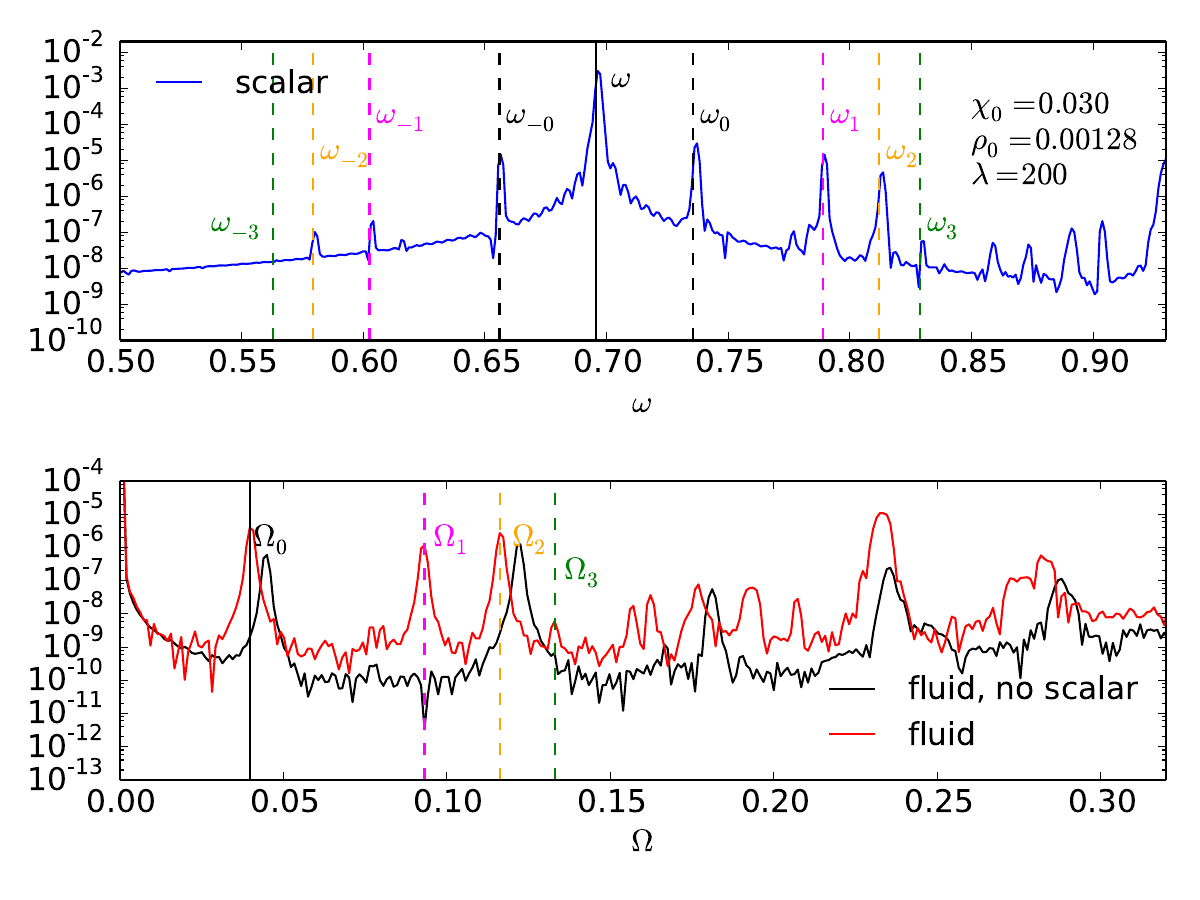}
        \par
	\end{centering}
	\caption{Fourier transforms of the amplitude of the real part of the scalar field and the central value of the fermion rest-mass density for $\chi_0=0.008$ (left panel) and $\chi_0=0.030$ (right panel), both for $\lambda=200$. }
\label{fig:freq1}
\end{figure}

Motivated by~\cite{lazarte2025gravitational}, we have investigated the oscillations and gravitational synchronization of the stable models in more detail. By performing a Fast Fourier Transform (FFT) of the time–series evolution of both the fluid central rest-mass and the scalar field amplitude, we can analyze the frequency content of the two components and identify how they are related. In Fig.~\eqref{fig:freq1}, we show the scalar-field and neutron star spectra for the mixed star with $\chi_0=0.008$ (top panel panel) and with $\chi_0=0.030$ (bottom panel), and $\lambda=200$. Focusing first on the scalar part (top panels), the black vertical line marks the oscillation frequency of the corresponding equilibrium configuration, which is the initial one. During the evolution, this remains the dominant mode, further proving the stability of the mixed configuration; however, several additional peaks of smaller amplitude appear, following a characteristic pattern that links the scalar-field oscillations to the radial modes of the neutron star.

In the bottom panels of Fig.~\eqref{fig:freq1}, we also display the frequencies extracted from the rest-mass density for both a pure neutron star (black line) and the neutron star in the mixed configuration. The neutron-star density oscillates approximately as

\begin{equation}
\rho(t,r) = \rho^{\rm static}(r)+\sum_i\rho^{i}\cos(\Omega_it),
\end{equation}
where $\Omega_0$ and $\Omega_2$ correspond to the fundamental mode and the first overtone of the pure neutron star, respectively~\cite{font2002three,Montero:2012yr,sanchis2014fully}. These same frequencies are present in the mixed case, but additional peaks also appear (labeled as $\Omega_1$ and $\Omega_3$), indicating the emergence of new oscillation modes induced by the coupling with the scalar field. The relations between the scalar and fluid frequencies are given by:
\begin{equation}
    \omega_{\pm i}=\omega\pm\Omega_i. \label{eq:cons_freq}
\end{equation}
These relations determined the first four peaks of the scalar field spectra, showing that we obtain a multi-frequency scalar configuration, which is synchronized with the radial oscillations of the neutron star, although dominated by the fundamental frequency $\omega_0$.

Interestingly, for a small scalar-field amplitude $\chi_0=0.008$, the oscillation spectrum of the neutron star closely resembles that of the fluid-only configuration. Although additional frequencies ($\Omega_1$ and $\Omega_3$) appear, their amplitudes are significantly smaller than those of the fundamental mode $\Omega_0$ and the first overtone $\Omega_2$. A similar behavior is observed in the scalar-field spectrum.

As $\chi_0$ increases, while the configuration remains stable, noticeable shifts in the neutron-star frequencies occur: both $\Omega_0$ and $\Omega_2$ are slightly displaced, and the amplitude of $\Omega_1$ becomes comparable to those dominant modes. These results indicate that gravitational synchronization in perturbed mixed configurations can give rise to richer and more complex oscillatory behavior.

\begin{figure}[H]
	\begin{centering}
		\includegraphics[scale=0.4]{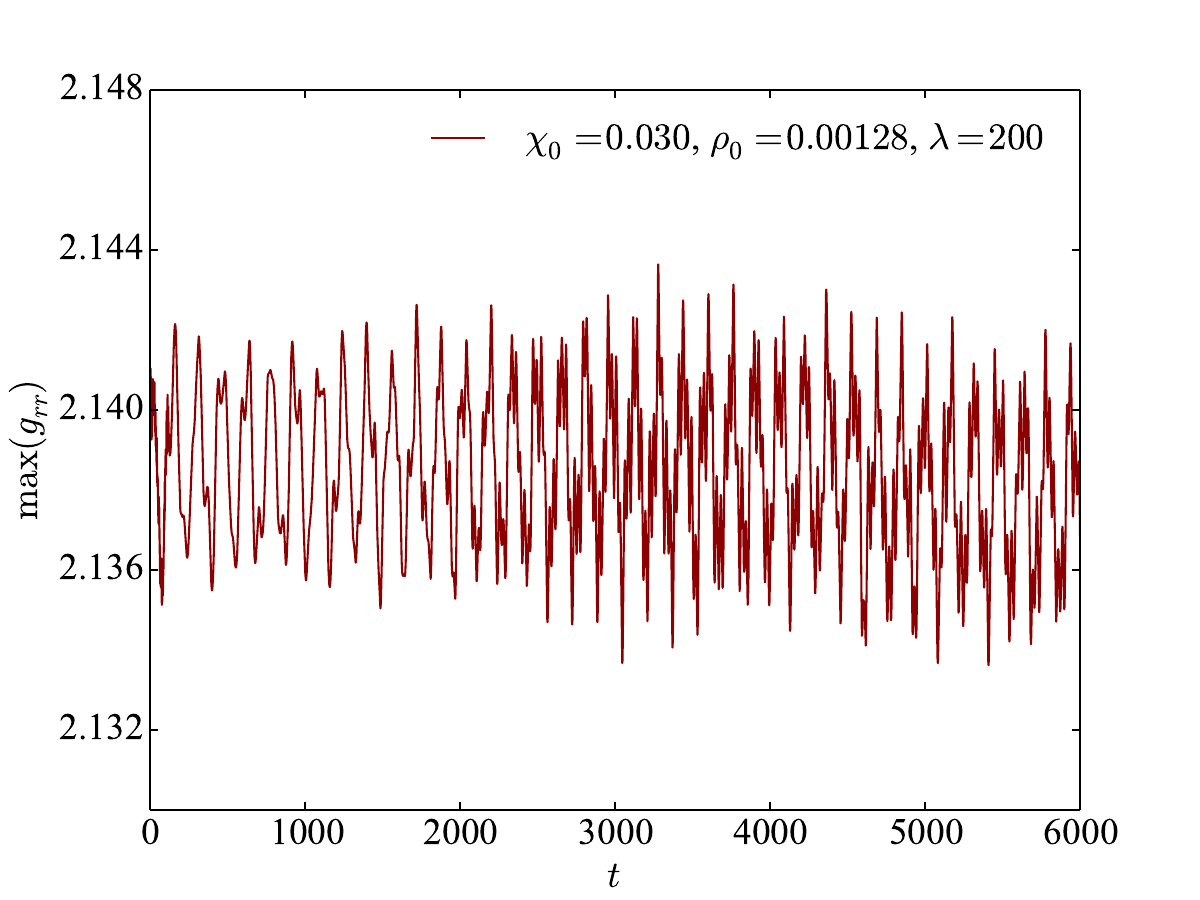}
        \includegraphics[scale=0.4]{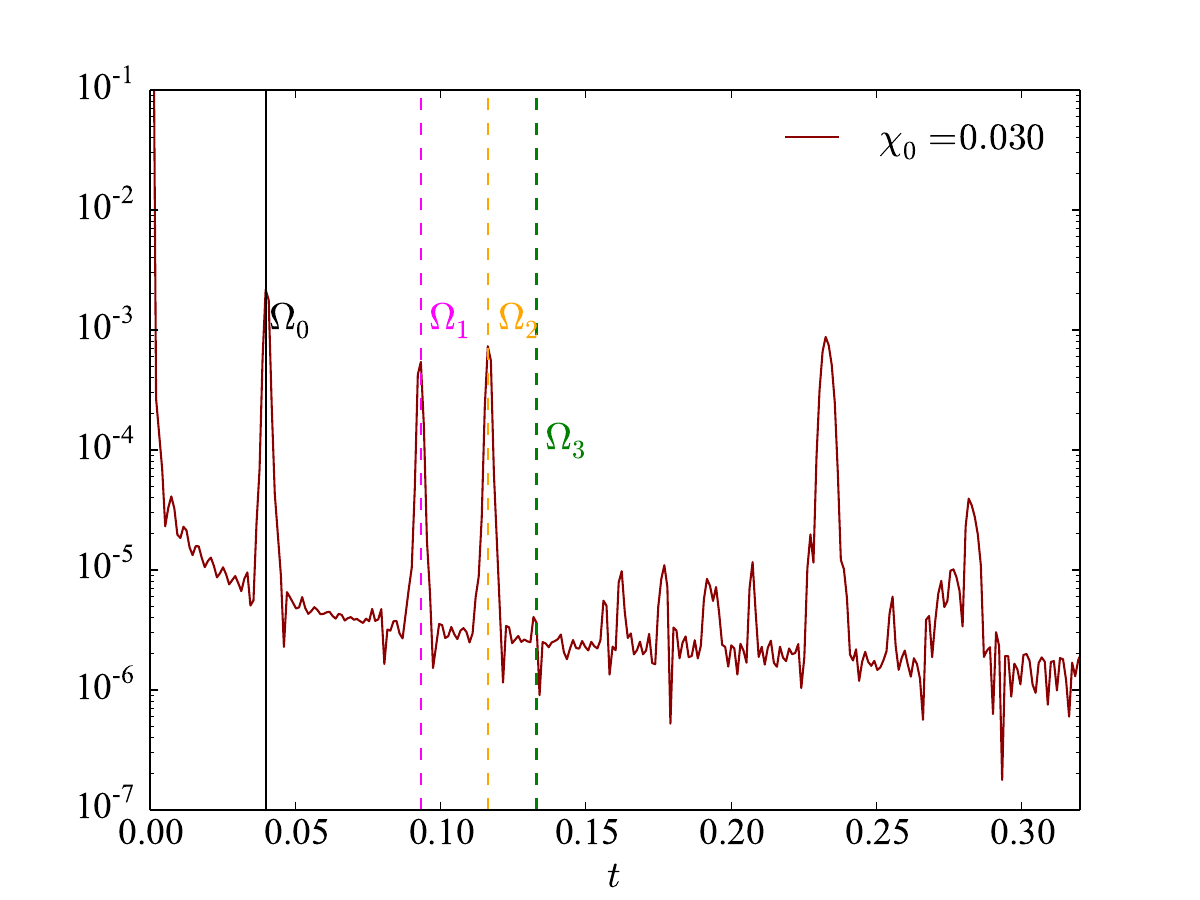}
        \par
	\end{centering}
	\caption{Time evolution of the maximum value of the metric component $g_{rr}$ and the Fourier transform for $\chi_0=0.030$ and for $\lambda=200$. }
\label{fig:freq2}
\end{figure}

Finally, in Fig.~\eqref{fig:freq2} we present the time evolution of the maximum value of the metric component $g_{rr}$, together with its Fourier transform, to verify that the frequencies match those of the fluid component. In the right panel of Fig.~\eqref{fig:freq2}, we identify the peaks in the spectrum as the neutron-star oscillation frequencies $\Omega_i$, thereby demonstrating that the synchronization occurs through gravitational coupling.

\section{Discussion} \label{sec:discussion}

Our results indicate that ghost scalar matter can be gravitationally confined within neutron stars. This is a nontrivial outcome: even in the extreme case of a field with a negative kinetic term (thereby violating the standard energy conditions) it is possible to construct regular, spatially bounded, and dynamically stable configurations. In this sense, our study shows that exotic matter sectors, including those that violate the null and weak energy conditions, cannot be categorically ruled out in strong gravitational fields at the classical level.

This finding suggests that compact objects embedded in a cosmological background containing dark-energy-like components could, in principle, accumulate and confine small fractions of such exotic matter. While realistic dark energy models do not necessarily require a ghost kinetic structure, our results indicate that even more extreme violations of the energy conditions can remain gravitationally localized in equilibrium configurations.

Interestingly, despite the negative kinetic term, the total energy of the system remains bounded in the obtained configurations after they are perturbed, admitting stable families of solutions for the coupled system. Contrary to the common expectation that ghost sectors inevitably lead to catastrophic behavior, our analysis shows that, when confined and treated at the classical level, they can participate in well-defined and dynamically stable compact objects. This provides a concrete framework in which the conceptual challenges associated with ghost matter can be explored through fully relativistic numerical experiments.

We also stress the oscillatory coupling  of the mixed stars. As the scalar field and the fermionic matter only interact by means of the gravitational interaction, such coupling is a very interesting phenomenon which has the spacetime as the mediator. Indeed, the oscillations of a given body, say the ghost field, produces oscillations in the metric coefficients which in turn makes the other body oscillate accordingly, satisfying a  
sort of frequency conservation, like the coupled oscillators mediated by a dynamical field, or the wave mixing phenomenon  seen in non linear optics, Eq.~\eqref{eq:cons_freq}.

Indeed, we find that the presence of a confined ghost core could induce a persistent pulsation of the host neutron star~\cite{lazarte2025gravitational}. This dynamical feature opens the possibility of observational signatures, as the modified oscillation spectrum could, in principle, be detectable. For specific choices of the scalar field parameters, namely the mass parameter $\mu$ and the self-interaction coupling $\lambda$, the amplitude and frequency of the induced pulsations can be quantitatively determined. A systematic exploration of the parameter space and its potential phenomenological implications will be presented in future work.

%%%%%%%%%%%%%%%%%%%%%%%%%%%
%%%   ACKNOWLEDGMENTS   %%%
%%%%%%%%%%%%%%%%%%%%%%%%%%%

\acknowledgments
%Argelia
This work was partially supported by SECIHTI, México, under grant CBF-2025-G-1720.
%Nestor
NAMH acknowledges the support of SECIHTI, México, through grant number 812415.
%Victor
VJ acknowledges support from the National Key R\&D Program of China under grant No.~2022YFC2204603 and from the Secretaría de Ciencia, Humanidades, Tecnología e Innovación (SECIHTI, Mexico) through a postdoctoral fellowship.
%Nico
NSG acknowledges support from the Spanish Ministry of Science, Innovation, and Universities via the Ram\'on y Cajal programme (grant RYC2022-037424-I), funded by  MICIU/AEI/10.13039/501100011033 and by ESF+. This work is further supported by the Spanish Agencia Estatal de Investigaci\'on (Grant PID2024-159689NB-C21) funded by  MICIU/AEI/10.13039/501100011033 and ERDF A way of making Europe,  and by the European Horizon Europe staff exchange (SE) programme HORIZON-MSCA2021-SE-01 Grant No. NewFunFiCO-101086251.

\appendix

\section{Convergence}

\begin{figure}[H]
	\begin{centering}
  \includegraphics[scale=.42]{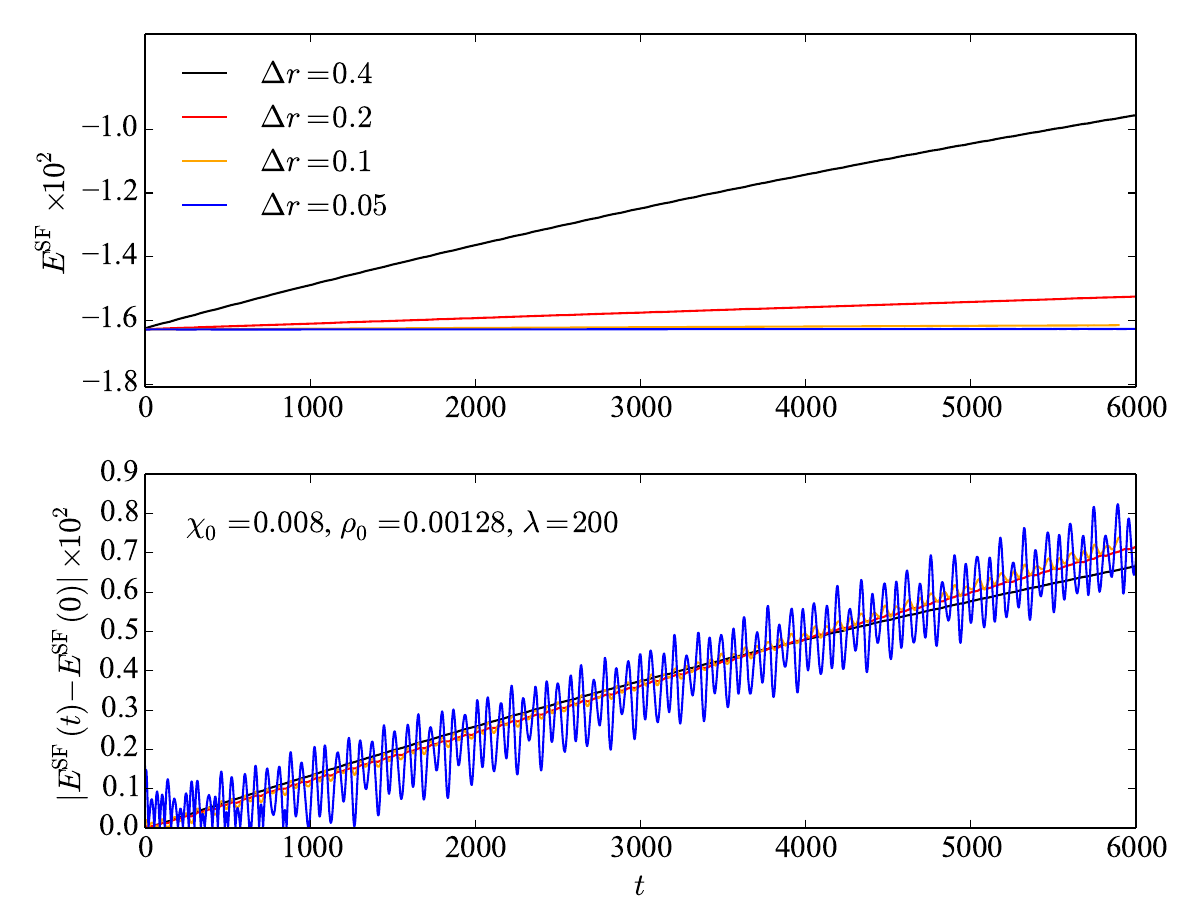}  \includegraphics[scale=.42]{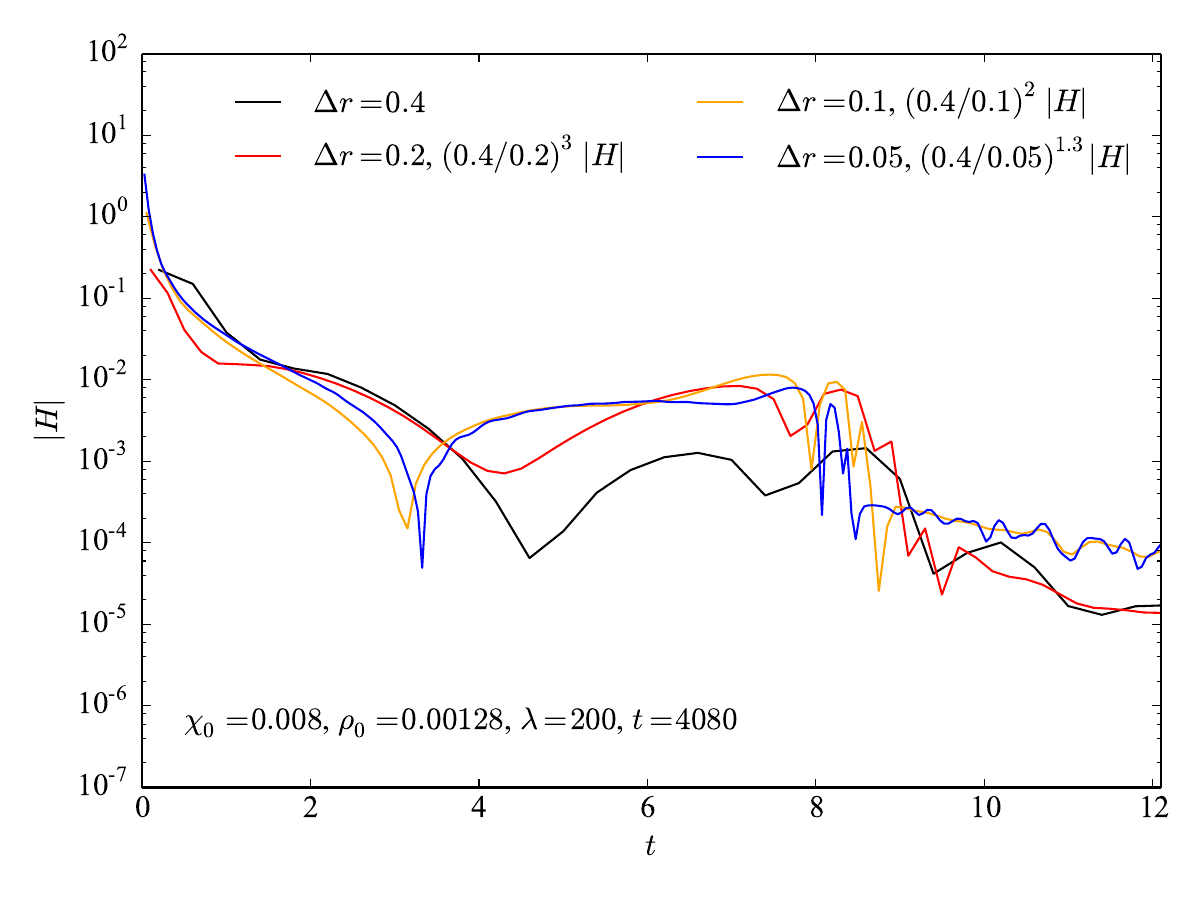}
		\par
	\end{centering}
	\caption{(Left panel) Time evolution of the real part of the scalar field extracted at $r_{\rm obs}=$ for different values of the initial $\chi_0=\lbrace0.0150,0.030\rbrace$ for the ghost boson star case. (Right panel) Time evolution of the central value of the lapse $\alpha$.}
\label{fig:conv1}
\end{figure}

We perform a convergence test comparing the evolution of different quantities using various grid resolutions. In the left panel of Fig.~\ref{fig:conv1}, we plot the evolution of the total scalar energy, as in Fig.~\ref{fig:energy_scalar}. The scalar mass should remain constant in time, as the model is stable $\chi_0=0.008$, $\rho_0=0.00128$, and $\lambda=200$. Due to numerical error, there is a drift in the mass and it decreases with time. Considering the deviation from the initial value of the mass $E^{\rm SF}(t=0)$, this quantity converges at third order.

In left panel of Fig.~\ref{fig:conv1} we show the absolute value of the Hamiltonian constraint at $t=4080$ for four different resolutions.
We find three to second-order convergence for the very low, low, and medium resolutions, ($\Delta r=0.4$, $\Delta r=0.2$, $\Delta r=0.1$). The convergence for higher resolution decreases to between first and second order. See~\cite{brito2023stability,brito2024self,lazarte2025gravitational} for more details.

\begin{figure}[H]
	\begin{centering}
		\includegraphics[scale=0.5]{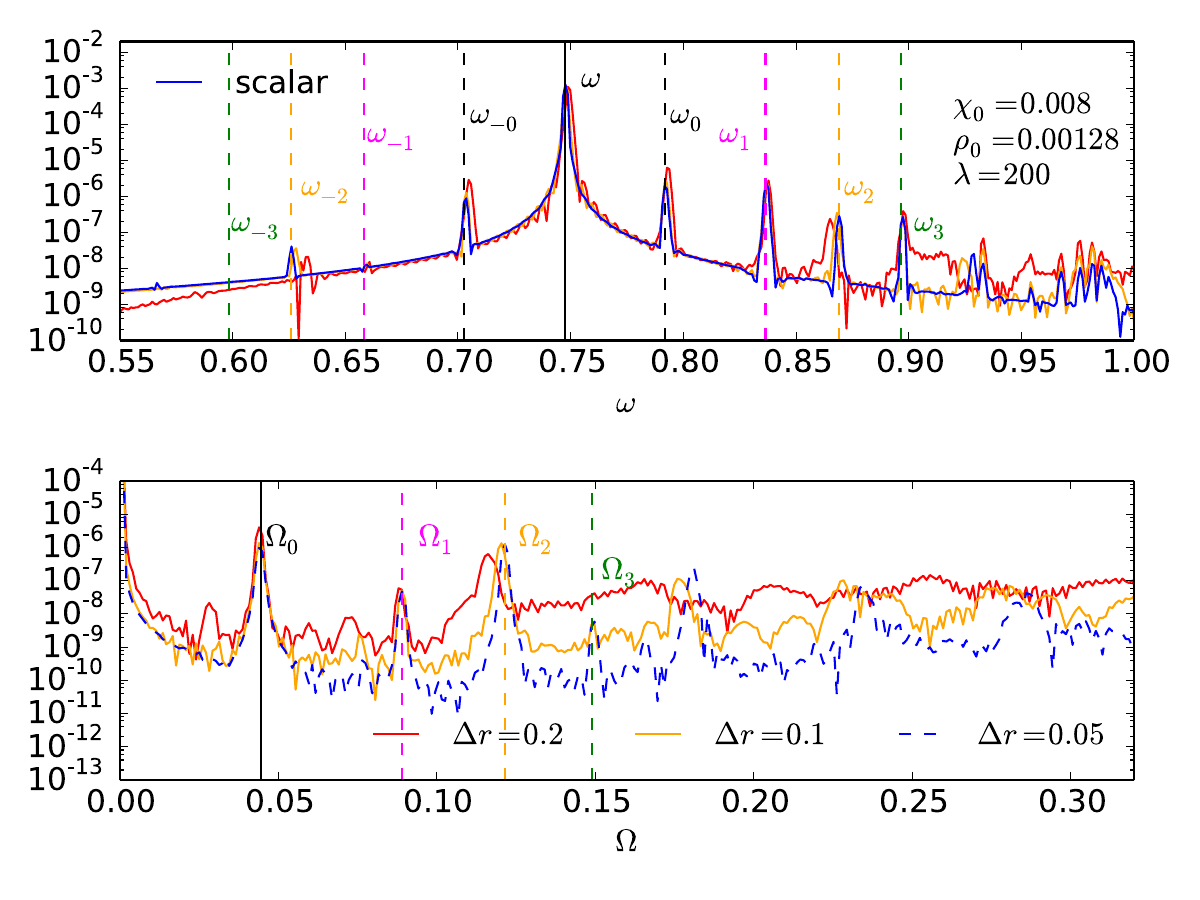}
        \par
	\end{centering}
	\caption{Fourier transforms of the amplitude of the real part of the scalar field and the central fermion rest-mass density for three different resolutions and $\chi_0=0.008$, $\lambda=200$. }
\label{fig:conv2}
\end{figure}

Finally in Fig.~\ref{fig:conv2} we plot the Fourier Transform of the scalar amplitude and the rest-mass density evolutions, showing the good agreement between resolutions.

%

%%%%%%%%%%%%%%%%%%%%%%%%%%%%%%%%%%%%%%
%\appendix

%%%%%%%%%%%%%%%%%%%%%% 
%%%   REFERENCES   %%%
%%%%%%%%%%%%%%%%%%%%%%

\bibliography{ref}
\end{document}